\documentclass[aps,11pt,superscriptaddress,longbibliography]{revtex4-1}
\usepackage{amsmath, amsfonts,amssymb}
\def\*#1{\mathbf{#1}}

\usepackage{float}
\usepackage{graphicx}
\usepackage{mathtools}
\DeclarePairedDelimiter\ceil{\lceil}{\rceil}

\usepackage{cleveref}
\newcommand{\quotes}[1]{``#1''}
\usepackage{comment}
\usepackage{color}
\usepackage{subfigure}

\begin{document}

\title{A simplified lattice Boltzmann implementation of the quasi-static \\   approximation in pipe flows under the presence \\  of non-uniform magnetic fields}
%\date{\today}
\author{H.S. Tavares}\email{hugoczpb@gmail.com.br}
\affiliation{Instituto de Física, Universidade Federal do Rio de Janeiro, C.P. 68528, CEP: 21945-970, Rio de Janeiro, RJ, Brazil}
\author{B. Magacho}%\email{}
\affiliation{Instituto de Física, Universidade Federal do Rio de Janeiro, C.P. 68528, CEP: 21945-970, Rio de Janeiro, RJ, Brazil}
\author{L. Moriconi}%\email{}
\affiliation{Instituto de Física, Universidade Federal do Rio de Janeiro, C.P. 68528, CEP: 21945-970, Rio de Janeiro, RJ, Brazil}
\author{J. Loureiro}%\email{}
\affiliation{Programa de Engenharia Mecânica,\\ Coordenação dos Programas de Pós-Graduação em Engenharia, \\ Universidade Federal do Rio de Janeiro,
C.P. 68503, CEP: 21945-970, Rio de Janeiro, RJ, Brazil}

    \begin{abstract}
    
    We propose a single-step simplified lattice Boltzmann algorithm capable of performing magnetohydrodynamic (MHD) flow simulations in pipes for very small values of magnetic Reynolds numbers $R_m$. In some previous works, most lattice Boltzmann simulations are performed with values of $R_m$ close to the Reynolds numbers for flows in simplified rectangular geometries. One of the reasons is the limitation of some traditional lattice Boltzmann algorithms in dealing with situations involving very small magnetic diffusion time scales associated with most industrial applications in MHD, which require the use of the so-called quasi-static (QS) approximation. Another reason is related to the significant dependence that many boundary conditions methods for lattice Boltzmann have on the relaxation time parameter. In this work, to overcome the mentioned limitations, we introduce an improved simplified algorithm for velocity and magnetic fields which is able to directly solve the equations of the QS approximation, among other systems, without preconditioning procedures. In these algorithms, the effects of solid insulating boundaries are included by using an improved explicit immersed boundary algorithm, whose accuracy is not affected by the values of $R_m$. Some validations with classic benchmarks and the analysis of the energy balance in examples including uniform and non-uniform magnetic fields are shown in this work. Furthermore, a progressive transition between the scenario described by the QS approximation and the MHD canonical equations in pipe flows is visualized by studying the evolution of the magnetic energy balance in examples with unsteady flows.

	\end{abstract}

\maketitle

 \tableofcontents

   \section{Introduction}
  
  Magnetohydrodynamics (MHD) flows are found in nature and in industrial applications involving many conductive fluids and plasma flows. In most of industrial applications, for example, the magnetic Reynolds number $R_m$ is very often smaller than $10^{-2}$~\cite{davidson2002introduction}. Simulations involving small values $R_m$ are usually performed by using the so-called quasi-static (QS) approximation, where the induced magnetic fluctuations are considered much smaller than the applied magnetic field~\cite{davidson2002introduction,knaepen2004magnetohydrodynamic,muller2001magnetofluiddynamics}. The derivation of the QS approximation involves taking the limit of vanishing $R_m$, which can introduce several challenges from the numerical point of view.  One of the biggest difficulties is associated with the need of solutions for a separate evolution equation for the magnetic field, and another difficulty comes with the presence of a very small diffusion time scale. Due to these difficulties, many numerical works in MHD have been restricted to cases where the magnetic Prandtl number $Pr_m$ is close to 1, i.e., where the magnetic and kinetic time scales are the same. This is also the case in many numerical works in the literature of the lattice Boltzmann methods (LBM)~\cite{de2021one,pattison2008progress,de2019universal,premnath2009steady}. In~\cite{pattison2008progress,premnath2009steady}, simulations with very small $Pr_m$ are performed but only in the context of stationary flows. 
  
  One of the main objectives in this article is to approach the equations of the QS regime by only using a lattice Boltzmann framework. More specifically, we aim to extend the simplified lattice Boltzmann models proposed in~\cite{delgado2021single,de2021one} for simulations of MHD flows involving curved boundaries with very small values of magnetic Reynolds numbers. In this analysis, we also intend to study the transition between the regime described by the canonical  MHD equations and the regime characteristic of the QS approximation~\cite{knaepen2004magnetohydrodynamic}. In our study, we manage to analyse not only the transition, but also regimes with $R_m \ll 1$, characteristic of industrial applications.
  
  In the original simplified single-step LBM~\cite{delgado2021single}, the straightforward introduction of the forcing terms does not take into consideration the lattice discrete effects, as pointed by~\cite{gao2021consistent,chen2018simplified} in some analogous simplified LBM models. Also, many simplified models have limitations with respect to the stability and accuracy for high values of relaxation times, the same limitation also appears in the classical LBM-BGK model \cite{kruger2017lattice,succi2018lattice}, which can be seen as one of the main limitations of this model towards simulations with small values of $R_m$. Another issue is associated with the dependence on the relaxation time parameter that some boundary conditions methods for LBM have, as pointed out by~\cite{gsell2019explicit}. The influence of curved boundaries was not addressed by~\cite{de2021one} in the context of MHD flows, and in the Ref.~\cite{pattison2008progress}, the only simulation involving curved boundaries is performed with $Pr_m=1$.
  
  By considering the recent advances provided by the works~\cite{gsell2019explicit,zhou2020macroscopic,gao2021consistent}, we manage to overcome many of the limitations of the previous lattice Boltzmann models by introducing an improved simplified LBM framework able to perform simulations of the QS approximation in flows with curved insulating boundaries up to $Pr_m \sim 10^{-7}$ in the laminar regime. Not only that, by considering preconditioning procedures~\cite{pattison2008progress,premnath2009steady,guo2004preconditioned,izquierdo2008preconditioned,turkel1999preconditioning}, we also manage to perform some simulations with $Pr_m>1$, a regime characterized by fast fluctuations of the magnetic fields, which require the use of more accurate numerical methods. In the LBM literature, to the best of our knowledge, only a few studies~\cite{de2018advanced,de2022vortex} analyzed MHD flows in this regime, showing accurate results up to $Pr_m=2$.

This article is organized as follows. In the first part, Section \ref{General equations and QS regime}, we describe the general MHD equations and its connections with the quasi-static approximations, enumerating some important differences between the two systems from the numerical point of view. In Section~\ref{Section about simplified LBM model}, we briefly introduce the traditional lattice Boltzmann method. In the following, we discuss a recent simplified single-step LBM algorithm for MHD flows based on the research developed by~\cite{de2021one,delgado2021single}. In the Section~\ref{Section of results and discussion}, we describe the general structure of the verification of benchmarks and validations considered throughout the article. In Section~\ref{Improved single step algorithm}, the single-step algorithm undergoes to a series of improvements, where increase of stability and accuracy are proposed with a some numerical validations. In the same section, a viscosity- and resistivity-independent immersed boundary method (IBM) able to simulate flows in the quasi-static regime is proposed. In Section~\ref{Effects of non-uniform magnetic fields}, we apply the improvements developed in the previous sections for MHD flows involving non-uniform magnetic fields. In Section~\ref{Prandtl bigger than 1}, techniques for the simulation of regimes with $Pr_m>1$ are developed with some numerical validations; and in Section~\ref{Conclusions and Perspectives}, we provide some conclusions and perspectives.

  \section{Magnetohydrodynamic equations and the quasi-static approximation}\label{General equations and QS regime}
  
\begin{figure}[H]
			\centering
			\includegraphics[scale=0.3]{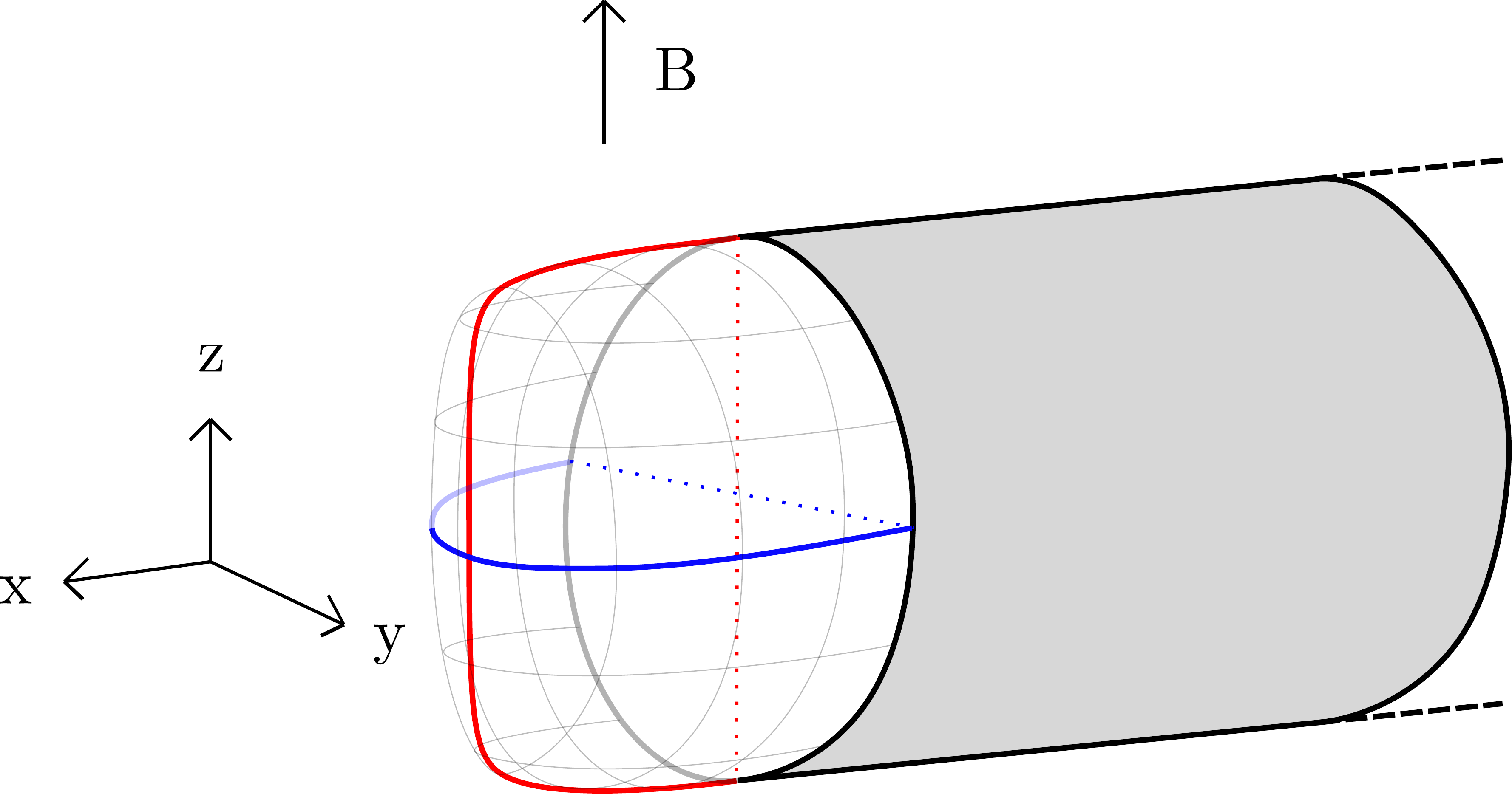}
			\caption{Schematic representation of MHD pipe flow with a transversal magnetic field. The red and blue lines correspond to some important cross sections of the velocity field that are analyzed in detail in this article. The analysis of these sections helps clarify what are the main effects of constant transverse magnetic fields.}
		\label{Schematic representation pipe flow}
	\end{figure}

  The equations describing magnetohydrodynamic phenomena are formed by a coupling between the continuity and the Navier–Stokes equations for describing the fluid motion, and the Maxwell's equations for electromagnetism as follows~\cite{davidson2002introduction}
	\begin{eqnarray}\label{complete system of MHD}
		\rho\left(\dfrac{\partial \*u }{\partial t}+(\*u \cdot \nabla)\*u \right) &=& -\nabla p  + \mu  \nabla^2 \*u + \*J \times \*B, \label{Navier-Stokes} \\
		\nabla \cdot {\bf u} &=&0, \label{Incompressibility-for-velocity-field}\\
		\dfrac{\partial \*B}{\partial t}+\nabla \cdot (\*u \otimes \*B-\*B\otimes\*u)&=&\eta \nabla^2 \*B, \label{ADE for magnetic field} \\
		\nabla \cdot \*B&=&0, \label{incompressibility magnetic field}
		\end{eqnarray}
	where $\*u$ and $\*B$ are the velocity and magnetic fields respectively, $\eta$ is the magnetic resistivity and $\mu$ is the dynamic viscosity of the fluid. We denote by $\nu=\mu/\rho$ the kinematic viscosity. For the sake of simplicity, in the rest of the article, we denote $\*u \otimes \*B=\*u \*B$ and $\*B \otimes \*u=\*B \*u$.  The electric field $\*E$ and the the electric current density $\*J$ are approximated by
	\begin{equation}\label{complete system quasi-static}
	 \*E=-(\*u\times \*B)+\eta (\nabla \times \*B), \ \ \ \ \*J=\nabla \times \*B.
	\end{equation} 
	Considering a system where $U_0$ is the characteristic velocity, $B_0$ is the characteristic magnetic intensity and L is the typical length scale. We have the following important dimensionless quantities
	\begin{equation}
	 \mathrm{Re}=\dfrac{U_0L}{\nu},  \ \ \ \  	\mathrm{R}_m=\dfrac{U_0L}{\eta},  \ \ \ \ \mathrm{Ha}= \dfrac{B_0L}{ \sqrt{\eta \nu}}, \ \ \ \ \ \mathrm{Pr}_m=\dfrac{\eta}{\nu},
	\end{equation}
	which are respectively: the Reynolds number, the magnetic Reynolds number, the Hartman number and the magnetic Prandtl number. In our study, we are mainly interested in the situations where $R_m\ll 1$, characteristic of the QS approximation~\cite{davidson2002introduction}, in pipe flows as shown schematically in Figure~\ref{Schematic representation pipe flow}. 
In this regime is convenient to introduce the decomposition $\*B=\*B^{ext}+\delta \*B$, where $\*B^{ext}$ is the external imposed magnetic field and $\delta \*B$ are fluctuations. The QS approximation translates into assuming \hbox{$\| \delta \*B \| \ll \| \*B^{ext} \|$}. 

 The following system holds in this regime~\cite{knaepen2004magnetohydrodynamic}
	\begin{eqnarray}\label{quasi-static approximation}
	\rho \left(	\dfrac{\partial \*u }{\partial t}+(\*u \cdot \nabla)\*u\right) &=& -\nabla\ p + \mu  \nabla^2 \*u + \*J \times \*B , \label{Navier-Stokes equations for QS} \\
		\nabla \cdot {\bf u} &=&0,\\
		\eta \nabla^2 \*B &=& \nabla \cdot (\*u  \*B^{ext}-\*B^{ext} \*u), \label{Poisson equation for QS} \\
		\nabla \cdot \*B&=&0.\label{zero divergence magnetic field}
		\end{eqnarray}

		This approximation does not involve the problems with very small magnetic diffusion time scales. The convection-diffusion equation (\ref{ADE for magnetic field}) for a magnetic field is replaced by a Poisson equation (\ref{Poisson equation for QS}). A first difficulty comes with these changes, which is the fact that usually the lattice Boltzmann  methods are not constructed to solve such types of equations. Also, in many problems, the solutions of Poisson equations involve non-local methods, which can be a problem if the objective is to perform parallelized simulations.  
		
		In the next sections, we aim to approach the system~(\ref{quasi-static approximation}-\ref{zero divergence magnetic field}) by using a lattice Boltzmann framework. In this approach, the problems with the very different diffusive time scales are handled by considering the asymptotic properties of a simplified LBM solver for advection-diffusion equations in order to treat the Poisson equation~(\ref{Poisson equation for QS}).
		The influence of curved walls is included by using an explicit immersed boundary method whose accuracy is not significantly affected by the coefficients of viscosity and resistivity. We also discuss lattice Boltzmann implementations of system 
		~(\ref{complete system of MHD}-\ref{incompressibility magnetic field}) for some simulations of pipe flows with $Pr_m>1$. 
		In the following sections, a detailed description of the described methods will be shown. 
	
	%\begin{itemize}
	%	\item The simplification eliminates the problem of a very {\bf small magnetic diffusion time scale} in the regime $Re_m<<1$;   
	%	\item But that brings up problems with the {\bf Poisson equation}.
	%	\item Some problems with traditional {\bf boundary conditions methods} for LBM.
	%\end{itemize}

	\section{Simplified single-step lattice Boltzmann methods for MHD flows}\label{Section about simplified LBM model}
	
	\subsection{Traditional lattice Boltzmann method}\hspace{0.3cm}
	
	The starting point of the lattice Boltzmann method is the connection between the Boltzmann equation and the classical hydrodynamics equations.  
	  The Boltzmann equation is an integro-differential equation for the probability density function $f(\*x,\*v,t)$ in six-dimensional space of a particle position $\*x\in \mathbb{R}^3$ and momentum $\*v\in \mathbb{R}^3$  given by
	\begin{equation}\label{Boltzmann equation original}
	\partial_{t}f+\nabla_{\*x}f\cdot \*v+\dfrac{\*F_{ext}}{\rho} \cdot \nabla_{\*v}f=Q(f,f),
	\end{equation}
	where $Q(f,f)$ is collision integral, $\*F_{ext}$  is the body force, $\rho$ is macroscopic mass density of the system, and $\nabla_{\*x}$ and $\nabla_{\*v}$ are gradients with respect to the position $\*x$ and velocity  $\*v$ coordinates, respectively. 	
	
	It can be shown that the collision integral $Q(f,f)$ has at least five invariants~\cite{wolf2004lattice}, i.e., a set of functions $\xi_k, \ k=1,2,3,4,5$, satisfying
	\begin{equation}
	\int_{\mathbb{R}^3} \xi_{k}(\*v) Q(f,f)d\*v=0,
	\end{equation} 
	which are $\xi_1=1$, $(\xi_2,\xi_3,\xi_4)=\*v$ and $\xi_5=|\*v|^2$. A general collision invariant can be written as linear combinations of the functions $\xi_{k}$. The invariants are associated to some important macroscopic quantities in the system, some of them are
	\begin{eqnarray}
	\textrm{mass density:} & & \int f d\*v=\rho, \label{density equation from Boltzmann} \\
	\textrm{momentum:}& & \int f \*v d\*v=\rho \*u.\label{Momentum equation from Boltzmann}
	\end{eqnarray}
	A set of conservation laws for each of these quantities can be obtained multiplying the Boltzmann equation (\ref{Boltzmann equation original}) by a collision invariant and subsequently integrating with respect to the velocity.
	
\begin{figure}[h!]
		\centering
		\includegraphics[scale=0.62]{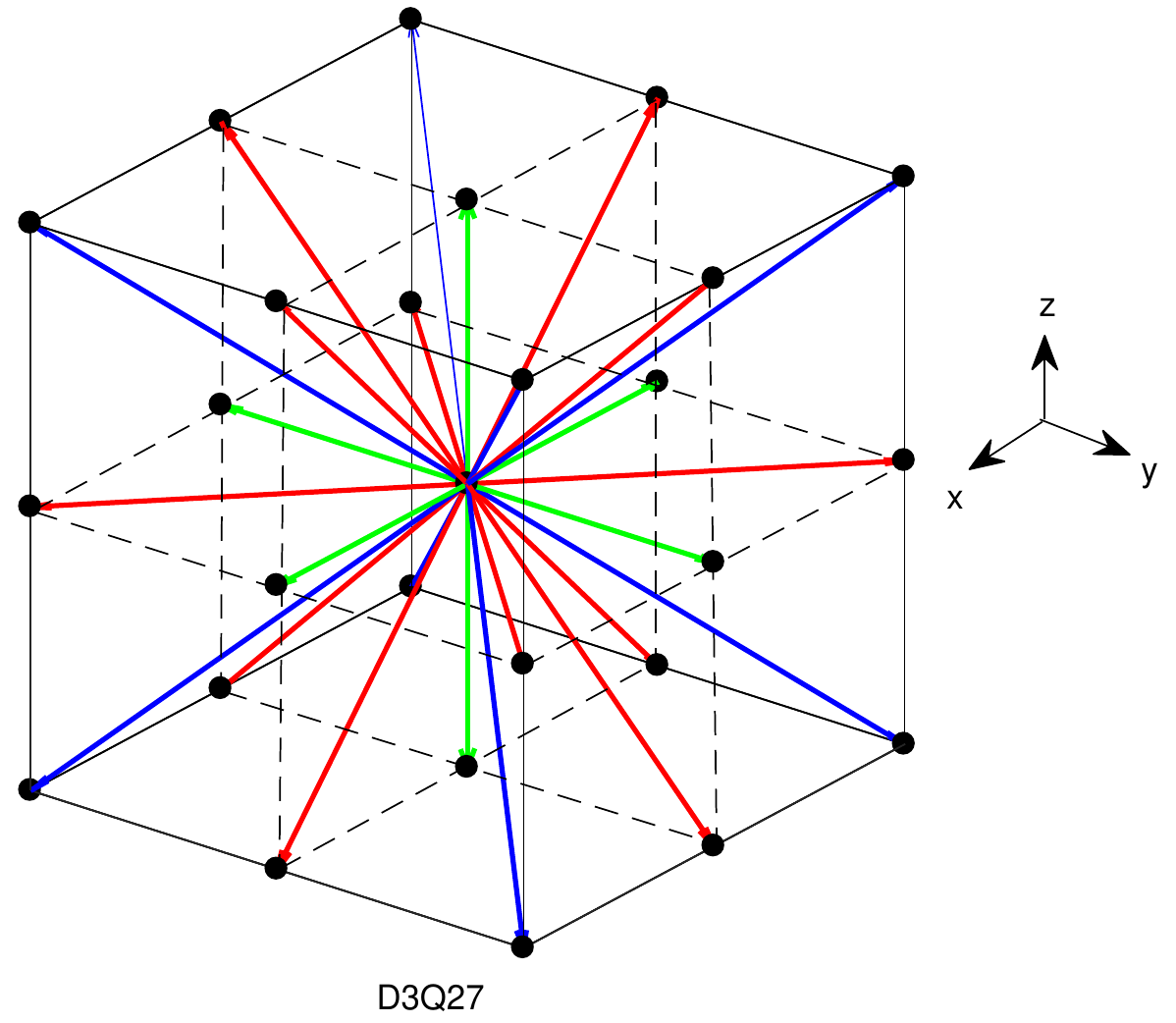}
		\caption{Lattice velocities for D3Q27 scheme.}
		\label{Lattice velocities D3Q27}
	\end{figure}

		\hspace{0.5cm}In the  lattice Boltzmann method (LBM) the basic quantity is the  discrete-velocity distribution function $f_i(\*x,t)$, it represents the density of particles with velocity $\*c_i$ at position $\*x$ and time $t$. By discretizing the Boltzmann equation (\ref{Boltzmann equation original}) in velocity space, physical space, and time, we obtain the discrete Boltzmann equation~\cite{kruger2017lattice,succi2018lattice}
	\begin{eqnarray}\label{General Kinematic Boltzmann equation}
	f_i(\*x+\*c_i \Delta t,t+\Delta t)=f_i(\*x,t)+\Omega_i(\*x,t),
	\end{eqnarray}
	where $\Omega_i(\*x,t)$ is the discrete version of the collision integral in~(\ref{Boltzmann equation original}).
	This equation expresses that a particle $f_i(\*x,t)$ moves with velocity $\*c_i$ to the nearest neighbors after a time step $\delta t$, i.e., the grid spacing is giving by $\delta x=|\*c_i|\delta t$. Analogously, the mass density and momentum density $\rho \*u$ at $(\*x,t)$ can be found through weighted sums known as moments of $f_i$ as
	\begin{eqnarray}
	\rho(\*x,t)&=&\sum_{i}^{}f_i(\*x,t),\label{Mass constraint} \\ 
	\rho(\*x,t) \*u(\*x,t)&=&\sum_{i}^{}\*c_i f_i(\*x,t), \label{Momentum constraint}
	\end{eqnarray}
	in a similar fashion to (\ref{density equation from Boltzmann}) and (\ref{Momentum equation from Boltzmann}). The main difference between $f_i$ and the continuous distribution function $f$ is that all of the argument variables of $f_i$ are discrete, with the subscript $i$ referring to a finite discrete set of velocities  {$\*c_i$} as shown in Figure~\ref{Lattice velocities D3Q27}.
	
		The discrete collision integral $\Omega_i$ is given by BGK operator defined as
	\begin{eqnarray}
	\Omega_i(f,f)=-\dfrac{f_i-f_i^{(eq)}}{\tau},
	\end{eqnarray}
	where the equilibrium distribution is given by
		\begin{eqnarray}\label{Equilibrium distribution simple fluids}
		f_i^{(eq)}(\rho,\mathbf{u})
		=w_i\rho\left(1+\dfrac{\mathbf{c}_i\cdot \mathbf{u}}{c_s^2}
		+\dfrac{(\mathbf{c}_i\cdot \mathbf{u})^2}{2 c_s^4}
		-\dfrac{\mathbf{u} \cdot \mathbf{u}}{2c_s^2}  \right),
		\end{eqnarray}
	where $c_s$ is the speed of sound given by $c_s=c/\sqrt{3}$ and $w_i$ are the lattice weights associated with the velocity scheme D3Q27 as shown in Table~I.
	\begin{table}[H]\label{Weights for the velocity scheme D3Q27}
	\centering
	\begin{tabular}{|l|c|c|c|}
	\hline
Velocities $\*c_i$ & Number & Weight $\textrm{w}_i$ \\
 \hline
(0,0,0)   & 1   & 8/27   \\
 \hline
($\pm 1$,0,0), \ (0,$\pm 1$,0), \ (0,0,$\pm 1$)    & 6      & 2/27  \\
 \hline
($\pm 1$,$\pm 1$,0), \ ($\pm 1$,0,$\pm 1$), \ (0,$\pm 1$,$\pm 1$)   & 12     & 1/54    \\
 \hline
 ($\pm 1$,$\pm 1$,$\pm 1$)   & 8   & 1/216    \\
 \hline
	\end{tabular}
\caption{Weights for the velocity scheme D3Q27.}
	\end{table}
	
		Using the BGK approximation in the equation (\ref{General Kinematic Boltzmann equation}), we obtain the lattice BGK equation
	\begin{eqnarray}\label{LBM original equation}
	f_i(\*x+\*c_i\delta t ,t+\delta t)=f_i(\*x,t)-\dfrac{1}{\tau}\left( f_i(\*x,t)-f_i^{(eq)}(\rho,\*u) \right). 
	\end{eqnarray}
	The simplest way to initialize the populations at the initial time $t=0$ is to set $f_i(\*x,t=0)=f^{(eq)}_i(\rho(\*x,t=0),\*u(\*x,t=0))$. The kinematic viscosity $\nu$ is connected to the relaxation time $\tau$ by the equation
	\begin{equation}
	    \nu = c_s^2\left( \tau - \dfrac{1}{2}    \right)\delta t.
	\end{equation}
	The BGK scheme is the most traditional LBM algorithm with many interesting applications, but it has well known limitations in terms of stability, memory
requirements and some problems with appropriate boundary conditions methods for some types of complex multiphysics
simulations~\cite{kruger2017lattice,succi2018lattice}.
	
    In the next section, we discuss  a recent approach that began with works developed by~\cite{delgado2021single,chen2017simplified,shu2014development}, later extended to MHD flows by~\cite{de2021one}, towards a simplified lattice Boltzmann method that does not involve the evolution of the non-equilibrium distributions. In this approach, a single-step algorithm is formulated giving a more efficient method in terms of memory requirements and stability in comparison with the traditional BKG algorithm~(\ref{LBM original equation}), while keeping almost the same accuracy.
	
		\subsection{Connection with hydrodynamic equations}\hspace{0.2cm}
	
	%Considering the long-wave–length and low-frequency limit, the time increment $\delta t$ and the grid spacing $\delta x=|\*c_i|\delta t$ in~(\ref{General Kinematic Boltzmann equation}) can be regarded as small parameters of order $\varepsilon$~\cite{chen1998lattice}. 
	
	From~(\ref{General Kinematic Boltzmann equation}), we can derive solutions for Navier-Stokes by first considering a 2nd-order Taylor series expansion in time and space given by
	\begin{equation}
	      f_i(\*x+\*c_i\delta t ,t+\delta t)-f_i(\*x,t)= \delta t D_if_i+ \dfrac{\delta t^2}{2} D_i^2f_i+O(\delta t^2),
	\end{equation}
	where $D_i=\frac{\partial}{\partial t}+\*c_i \cdot \nabla$ denotes the material derivative. Up to a second order error, we have~\cite{chen1998lattice}
	\begin{equation}\label{Taylor-expansion of kinetic algorithm}
	\frac{\partial}{\partial t}f_i+\*c_i \cdot \nabla f_i+\delta t \left( \dfrac{1}{2} \*c_1 \otimes \*c_i : \nabla \nabla f_i+\*c_i \cdot \nabla \dfrac{\partial f_i}{\partial t} +\dfrac{1}{2} \dfrac{\partial^2 f_i }{\partial t^2}   \right)=- \dfrac{1}{\tau \delta t} \left( f_i - f_i^{eq} \right).
	\end{equation}
	Next, consider the Chapman-Enskog multiscale expansion~\cite{frisch1995turbulence},
	\begin{equation}\label{Chapmann-Enkog multiscale}
	    \dfrac{\partial}{\partial t} = \varepsilon \dfrac{\partial}{\partial t_1}+ \varepsilon^2 \dfrac{\partial}{\partial t_2}, \ \ \ \ \ 
	    \nabla = \varepsilon \nabla_1.
	\end{equation}
	where $\varepsilon$ is a small parameter proportional to the Knudsen number~\cite{succi2018lattice}. 
	In this expansion, it is assumed that the diffusion time scale $t_2$ is much larger than the convective time scale $t_1$, and that diffusion and convection act on the same spatial scale~\cite{wolf2004lattice}. In  similar fashion, the distribution function $f_i$ can be expanded about the local equilibrium distribution function $f_i^{eq}$ as
	\begin{equation}\label{Expansion of the equilibrium distribution}
	    f_i = f_i^{eq}+\varepsilon f_i^{neq},
	\end{equation}
	where $f_i^{neq}=f_i^{(1)}+\varepsilon f_i^{(2)}+O(\varepsilon^2)$ is the nonequbilibrium distribution, which is associated with viscous dissipation and verifies the following constraints
	\begin{equation}\label{constraints for non-equilibrium}
	    \sum_{i}f^{(k)}_i = 0, \ \ \ \ \sum_{i}f^{(k)}_i \*c_i =1, \ \ \ \textrm{for} \ \ \ k=1,2,
	\end{equation}
	called solvability conditions.
 Substituting (\ref{Chapmann-Enkog multiscale}) and (\ref{Expansion of the equilibrium distribution})  into (\ref{Taylor-expansion of kinetic algorithm}) and combining the sequence of equations obtained up to order $O(\varepsilon^2)$, we obtain the following system~\cite{kruger2017lattice,succi2018lattice}
	\begin{equation}\label{first equation Chapmann}
\sum_{i=0}^N \left[ \dfrac{\partial f_{i}^{eq}}{\partial t}+ \*c_i \cdot \nabla f_i^{eq}\right] =0,
	\end{equation}
	\begin{equation}\label{second equation Chapmann}
\sum_{i=1}^{N} \*c_i \left[  \dfrac{\partial f_i^{eq}}{\partial t} + \*c_i \cdot \nabla f_i^{eq}+\left( 1-\dfrac{1}{2\tau} \right)  D_if_i^{(1)}   \right]=0,
	\end{equation}
	with
	\begin{equation}\label{non-equilibrium term}
	    f_i^{(1)}({\bf x},t) \simeq -\tau \delta t D_i f_i^{eq}(\*x,t) = -\tau \delta t \left(  \dfrac{\partial }{\partial t}+\*c_i \cdot \nabla   \right)f_i^{eq}(\*x,t).
	\end{equation}
	
By using the moments~(\ref{Mass constraint}), (\ref{Momentum constraint}) and (\ref{constraints for non-equilibrium}), it follows that the equations (\ref{first equation Chapmann}) and (\ref{second equation Chapmann}) can be turned into solutions for continuity and Navier-Stokes equations respectively~\cite{kruger2017lattice}. 
	
	\subsection{Single-step lattice Boltzmann algorithm for the Navier-Stokes equations}\hspace{0.2cm}
	
	The equations (\ref{first equation Chapmann}) and (\ref{second equation Chapmann}) 
	are the starting point of many simplified LBM algorithms~\cite{chen2017simplified,shu2014development,delgado2021single}. Different discretization schemes for these equations produce different simplified algorithms. In this article, the starting point is the approach developed by~\cite{delgado2021single}, which will be described in the following with a slightly different derivation.

   Considering the finite differences
	\begin{eqnarray}
	\dfrac{\partial f_{\alpha}^{eq}}{\partial t}&=& \dfrac{f_{\alpha}^{eq}({\bf x},t+\delta t)-f_{\alpha}^{eq}({\bf x},t)}{\delta t},\\
	{\bf c}_{\alpha} \cdot \nabla f_{\alpha}^{eq}&=& - \dfrac{f_i^{eq}(\*x+\*c_i\delta t,t)-4f_i^{eq}(\*x,t)+3f_i^{eq}(\*x-\*c_i\delta t,t) }{2\delta x},
\end{eqnarray}
we can rewrite (\ref{first equation Chapmann}) as
	\begin{eqnarray}\label{Equation 4}
	\sum_{i=1}^{N}\dfrac{f_i^{eq}(\*x,t+\delta t)-f_i^{eq}(\*x,t)}{\delta t}-\dfrac{f_i^{eq}(\*x+c_i\delta t,t)-4f_i^{eq}(\*x,t)+3f_i^{eq}(\*x-\*c_i\delta t,t) }{2\delta x}=0.
	\end{eqnarray}
	Using (\ref{Mass constraint}), we arrive in the following algorithm
	\begin{equation}\label{Single-step for density equation}
\rho(\*x,t+\delta t)= \dfrac{3}{2}\sum_{i=1}^{N}f^{eq}(\*x-\*c_i \delta t,t)-\sum_{i=1}^{N}f^{eq}_i(\*x,t) +\dfrac{1}{2}\sum_{i=1}^{N} f_i^{eq}(\*x+\*c_i \delta t,t).
\end{equation}

For the momentum equation (\ref{second equation Chapmann}), the term $\*c_i \cdot \nabla f_i^{eq}$ is discretized in a different way as
\begin{eqnarray}\label{Gradient discretization for momentum}
     \*c_i \cdot \nabla f_i^{eq}=\dfrac{  f_i^{eq}(\*x+\*c_i \delta t,t)-f_i^{eq}(\*x-\*c_i \delta t,t) }{2 \delta x }.
\end{eqnarray}
For the non-equilibrium term (\ref{non-equilibrium term}), we apply the directional approach for the gradient operation
	\begin{eqnarray}
	\sum_{i=1}^{N} \*c_i D_if_i^{(1)} = \sum_{i=1}^{N} \*c_i (\*c_i \cdot \nabla f_i^{(1)} ) =\sum_{i=1}^{N} \*c_i \dfrac{\partial f_i^{(1)}}{\partial \*c_i},
	\end{eqnarray}
	where we used the constraints in (\ref{constraints for non-equilibrium}). Using (\ref{non-equilibrium term}), we have
	\begin{eqnarray}\label{Approximation of time derivative of non-equilibrium}
	             \sum_{i=1}^{N} \*c_i \dfrac{\partial f_i^{(1)}}{\partial \*c_i} &\simeq & \sum_{i=1}^{N} \*c_i \dfrac{\partial }{\partial \*c_i} \left( -\tau \delta t \dfrac{\partial f^{eq}_i }{\partial \*c_i}  \right),\nonumber \\
	             &\simeq &\sum_{i=1}^{N} -\tau \delta t \*c_i \dfrac{\partial }{\partial \*c_i} \left(  \dfrac{  f_i^{eq}(\*x+\*c_i \delta t,t)-f_i^{eq}(\*x,t) }{ \delta x}  \right),\nonumber \\
	             &\simeq &\sum_{i=1}^{N} -\tau \delta t \*c_i  \left(  \dfrac{f_i^{eq}(\*x+\*c_i \delta t,t)-2 f_i^{eq}(\*x,t)+f_i^{eq}(\*x-\*c_i \delta t,t)}{(\delta x)^2}   \right), 
	\end{eqnarray}
	where we combined forwards and backwards finite differences for the operator $\frac{\partial}{\partial \*c_i}$.
	
	Substituting (\ref{Equation 4}), (\ref{Gradient discretization for momentum}) and(\ref{Approximation of time derivative of non-equilibrium})  into (\ref{second equation Chapmann}), and considering (\ref{Momentum constraint}) it follows that
	\begin{eqnarray}\label{Single-step for momentum equation}
	\rho(\*x,t+\delta t) \*u(\*x,t+\delta t)&=&  \sum_{i=1}^{N} \left\lbrace  \*c_i  f^{eq}_i(\*x-\*c_i \delta t,t)+\right. \\
	&+& \left. (\tau -1)\*c_i[f^{eq}_i(\*x+\*c_i \delta t,t)-2f^{eq}_i(\*x,t)+ f^{eq}_i(\*x-\*c_i \delta t,t) ] \right\rbrace.\nonumber
	\end{eqnarray}
  The equations (\ref{Single-step for density equation}) and (\ref{Single-step for momentum equation})  constitute the  single-step lattice Boltzmann algorithm~\cite{delgado2021single}. It is important to observe that these formulas depend only on the equilibrium distributions, which are only associated with the macroscopic quantities of the system. This feature reduces significantly the memory requirements in comparison to the traditional BGK algorithm, and also simplifies the implementation of boundary conditions, as we no longer have to deal with complicated manipulations of non-equilibrium distributions at the boundaries. In the next section, we consider a similar development in the context of the advection-diffusion equation~(\ref{ADE for magnetic field}) for the canonical MHD system.
  
  \subsection{Single-step simplified LBM algorithm for the magnetic fields equations}\hspace{0.2cm}
  
  In~\cite{dellar2002lattice}, Dellar derived an extension of the lattice BKG scheme~(\ref{LBM original equation}) that solves the advection-diffusion equation~(\ref{ADE for magnetic field}) for the magnetic field. This work also presents, in a similar fashion, the following algorithm
	\begin{eqnarray}\label{original LBM for magnetic field}
	g_{ix}(\*x+\*c_i \delta t,t+\delta t)= g_{ix}(\*x,t)-\dfrac{1}{\tau_m} (g_{ix}(\*x,t)-g_{ix}^{eq}(\*x,t)),
	\end{eqnarray}
	which solves, for example, the x-components of the magnetic field as 
 \begin{equation}
     B_x(\*x,t)=\sum_{i=1}^{N} g_{ix}(\*x,t).
 \end{equation}
 The relationship between resistivity $\eta$ and the relaxation parameter $\tau_m$ is given by
	\begin{eqnarray}
	\eta=c_s^2 \left(\tau_m -\dfrac{1}{2}  \right),
	\end{eqnarray}
where $c_s$ is the corresponding speed of sound. An
analogous equilibrium distribution is defined as
\begin{equation}\label{Equilibrium distribution for magnetic field}
	g_{ix}^{eq}(\*x,t) = w_i \left[ B_x + \dfrac{c_{iy}}{c_s^2} (u_y B_x - u_x B_y)+ \dfrac{c_{iz}}{c_s^2} ( u_z B_x - u_x B_z) )   \right].
\end{equation}

In the work~\cite{de2021one}, the authors introduced a single-step (or one-stage) simplified LBM algorithm for (\ref{ADE for magnetic field}) following the same steps of~\cite{delgado2021single}, as we describe as follows.

  	\hspace{0.5cm}The lattice Boltzmann equation (LBE) can be written as
	\begin{eqnarray}\label{LBM equation for magnetic equation}
	g_{ix}(\mathbf{x}+c_i \delta t)-g_{i}(\*x,t)=\dfrac{g_{ix}^{eq}(\*x,t)-g_{ix}(\*x,t)}{\tau_m}.
	\end{eqnarray} 
	By applying a Taylor series expansion at the left-hand side of (\ref{LBM equation for magnetic equation})
	followed by a Chapman–Enskog expansion up to second order, it is possible to write the following equation
	\begin{eqnarray}
	\sum_{i} \left[  \dfrac{\partial g_{ix}^{eq}}{\partial t}+ c_i\cdot\nabla g_{ix}^{eq}+ \left(1-\dfrac{1}{2\tau_{m}}\right) D_i g_{ix}^{(1)}\right]=0,
	\end{eqnarray}
	with
	\begin{equation}
	    g^{(1)}(\*x,t)\simeq - \tau_m \delta t D_ig_i^{eq} = - \tau_m \delta t \left(  \dfrac{\partial}{\partial t} + \*c_i \cdot \nabla   \right) g_i^{eq}(\*x,t).
	\end{equation}
	Now consider the following finite differences schemes
\begin{eqnarray}
\dfrac{\partial g_{ix}^{(0)}}{\partial t}&=&
\dfrac{g_{ix}^{(0)}(\*x+\*c_i \Delta t,t)-g_{ix}^{(0)}(\*x,t)}{\delta t},\\
\*c_i \cdot \nabla g_{ix}^{(0)}&=& \dfrac{g_{ix}^{(0)}(\*x+\*c_i \delta t,t)-g_{ix}^{(0)}(\*x-\*c_i \delta t,t)}{2\delta x},
\end{eqnarray}
and
\begin{eqnarray}
           \sum_{i=1}^{N} \*c_i  D_i g_{ix}^{(1)}&=& \sum_{i=1}^{N}  \*c_i(\*c_i \cdot \nabla g^{(1)}_{ix})\simeq\sum_{i=1}^{N}  \*c_i\dfrac{g_{ix}^{(1)}(\*x+\*c_i \delta t,t)-g_{ix}^{(1)}(\*x,t)}{\delta x}\simeq \nonumber \\
           &\simeq& \sum_{i=1}^{N} - \*c_i  \tau_m \delta t \dfrac{\left[ g_{ix}^{eq}(\*x+\*c_i \delta t,t)-2 g_{ix}^{eq}(\*x,t)+g_{ix}^{eq}(\*x-\*c_i \delta t,t) \right]}{(\delta x)^2}.
\end{eqnarray}
So it follows that
	\begin{eqnarray}
		\sum_i \left[  g_{ix}^{eq}(\*x,t+\delta t)+2 (\tau_{m} -1 ) g_{ix}^{eq}(\*x,t)-\right. \nonumber \\
	\left. -(\tau_m-1) g_{ix}^{eq}(\*x+\*c_i \delta t, t)-\tau_{\eta} g_{ix}^{eq}(\*x-\*c_i \delta t,t)\right]=0,
	\end{eqnarray}
	and then,
		\begin{eqnarray}\label{single-step algorithm for MHD de rosis}
		B_x(\*x,t+\delta t)&=&\sum_{i=1}^{N} \left\lbrace g_{ix}^{eq} (\*x-\*c_i \delta t,t) + \right. \\
		&+& \left. (\tau_m-1)[g_{ix}^{eq} (\*x+\*c_i \delta t,t)-2g_{ix}^{eq} (\*x,t)+g_{ix}^{eq} (\*x-\*c_i \delta t,t)] \right\rbrace.\nonumber
		\end{eqnarray}
		Analogously, the algorithm is only a function of the equilibrium distribution given by~(\ref{Equilibrium distribution for magnetic field}). This algorithm is also usually much more stable then the traditional form~(\ref{original LBM for magnetic field}).
		
		\subsection{Summary of the one-stage simplified LBM algorithm for MDH flows}\hspace{0.2cm}
		
			\hspace{0.5cm}	Considering the following expressions for the equilibrium distributions:
		\begin{eqnarray}
		f_i^{eq}(\*x,t)
		&=&w_i\rho\left(1+\dfrac{\mathbf{c}_i\cdot \mathbf{u}}{c_s^2}
		+\dfrac{(\mathbf{c}_i\cdot \mathbf{u})^2}{2 c_s^4}
		-\dfrac{\mathbf{u} \cdot \mathbf{u}}{2c_s^2} \right),\\
		g_{ix}^{eq}(\*x,t) &=& w_i \left[ B_x + \dfrac{c_{iy}}{c_s^2} (u_y B_x - u_x B_y)+ \dfrac{c_{iz}}{c_s^2} ( u_z B_x - u_x B_z) )   \right],\label{Equilibrium distribution for magnetic field I} \\
		g_{iy}^{eq}(\*x,t) &=& w_i \left[ B_y + \dfrac{c_{ix}}{c_s^2} (u_x B_y - u_y B_x)+ \dfrac{c_{iz}}{c_s^2} ( u_z B_y - u_y B_z) )   \right],\label{Equilibrium distribution for magnetic field II}\\
		g_{iz}^{eq}(\*x,t) &=& w_i \left[ B_z + \dfrac{c_{ix}}{c_s^2} (u_x B_z - u_z B_x)+ \dfrac{c_{iy}}{c_s^2} ( u_y B_z - u_z B_y) )   \right].  \label{Equilibrium distribution for magnetic field III}
		\end{eqnarray}

	\hspace{0.5cm}	We have the following single-step (or one-stage) LBM algorithm for MHD flows
	\begin{eqnarray}
	\rho(\*x,t+\delta t)&=& \sum_{i=1}^{N} \dfrac{3}{2} f^{eq}(\*x-\*c_i \delta t,t)-f^{eq}_i(\*x,t) +\dfrac{1}{2}f_i^{eq}(\*x+\*c_i \delta t,t),\nonumber \\
\rho(\*x,t+\delta t)	\*u(\*x,t+\delta t)&=&\sum_{i=1}^{N} \*c_i \left\lbrace f^{eq}_i(\*x-\*c_i \delta t,t)+\right. \label{single-step algorithm complete}\\
	&+& \left. (\tau -1)[f^{eq}_i(\*x+\*c_i \delta t,t)-2f^{eq}_i(\*x,t)+ f^{eq}_i(\*x-\*c_i \delta t,t) ]\right\rbrace,\nonumber\\
	\*B(\*x,t+\Delta t)&=&\sum_{i=1}^{N} \left\lbrace \*g_{i}^{eq} (\*x-\*c_i \delta t,t) + \right. \nonumber \\ 
	&+& \left. (\tau_m-1)[\*g_{i}^{eq} (\*x+\*c_i \delta t,t)-2\*g_{i}^{eq} (\*x,t)+\*g_{i}^{eq} (\*x-\*c_i \delta t,t)] \right\rbrace,\nonumber
	\end{eqnarray}
	where $\*B=[B_x,B_y,B_z]$ and $\*g_i^{eq}=[g_{ix}^{eq},g_{iy}^{eq},g_{iz}^{eq}]$.
	External forcing terms $\*F_{ext}$ are usually added in a straightforward way as
	\begin{eqnarray}\label{Single step forcing term introdution}
	 \rho(\*x,t+\delta t)   \*u(\*x,t+\delta t)&=&\sum_{i=1}^{N} \*c_i \left\lbrace f^{eq}_i(\*x-\*c_i \delta t,t)+ \right. \nonumber\\
	&+& \left. (\tau -1)[f^{eq}_i(\*x+\*c_i \delta t,t)-2f^{eq}_i(\*x,t)+ f^{eq}_i(\*x-\*c_i \delta t,t)]\right\rbrace+ \\
	&+&\*F_{ext} \delta t. \nonumber
	\end{eqnarray}
	 Dirichlet boundary conditions for geometries formed by flat boundaries are implemented straightforward by just assigning the desired values to the boundary points, some other types of boundary conditions are also implemented very similarly to
conventional MHD solvers. To the best of our knowledge, no studies of the single-step LBM algorithm have been conducted in the context of MHD flows involving curved boundaries. The success of the use of the immersed boundary methods~\cite{kruger2017lattice} in some previous lattice Boltzmann models~\cite{chen2017simplified,gsell2019explicit,zhao2021efficient} indicates an interesting direction for the inclusion of curved boundaries in simulations of MHD flows.
	
	%In order to minimize the influence of acoustic waves and to improve the accuracy, in this article we modify the equilibrium distribution in the following way~\cite{he1997lattice}
	%\begin{equation}
	 %   f_i^{eq}(\*x,t)
	%	=w_i\left[\dfrac{p}{c_s^2}+\left(\dfrac{\mathbf{c}_i\cdot \mathbf{u}}{c_s^2}
	%	+\dfrac{(\mathbf{c}_i\cdot \mathbf{u})^2}{2 c_s^4}
	%	-\dfrac{\mathbf{u} \cdot \mathbf{u}}{2c_s^2} \right) \right].
	%\end{equation}
	%As a consequence, the one-stage algorithm for the Navier-Stokes equation is modified as
	%\begin{eqnarray}
	% \dfrac{1}{c_s^2} p(\*x,t+\delta t)&=& \sum_{i=1}^{N} \dfrac{3}{2} f^{eq}(\*x-\*c_i \delta t,t)-f^{eq}_i(\*x,t) +\dfrac{1}{2}f_i^{eq}(\*x+\*c_i \delta t,t),\label{single-step densities} \\
	%\*u(\*x,t+\delta t)&=&\sum_{i=1}^{N} \*c_i  f^{eq}_i(\*x-\*c_i \delta t,t)+ \label{single-step momentum}\\
	%&+& (\tau -1)\*c_i[f^{eq}_i(\*x+\*c_i \delta t,t)-2f^{eq}_i(\*x,t)+ f^{eq}_i(\*x-\*c_i \delta t,t) ]+{\bf F_{ext}} \delta t,\nonumber
	%\end{eqnarray}

	It important to observe that  the inclusion of the forcing terms by using (\ref{Single step forcing term introdution}) does not consider the so-called lattice discrete effects~\cite{gao2021consistent}, associated to the correct consideration of contribution of the fording term $\frac{\*F_{ext}}{\rho} \cdot \nabla_{\*v} f$ in the equation~(\ref{Boltzmann equation original}). This limitation can compromise the accuracy of the simulations, especially in the cases involving non-uniform or unsteady forcing terms.   
	
	Another limitation of the algorithm~(\ref{single-step algorithm complete}) is associated with the loss of stability and accuracy for high values of relaxation times. Considering $\delta t=\delta x=1$, the simulations become easily unstable for values of relaxation times $\tau>0.5$ and $\tau_m>0.5$, a similar limitation is also shared by other simplified methods. In our work, one of the main objectives is the to solve the quasi-static approximation in MHD, and for this objective is necessary to consider high values of resistivity $\eta$ which usually implies in very high values of $\tau_m$. 
	
	In the next sections, we address all of the mentioned limitations. We first consider an implementation of forcing scheme that takes into consideration the effects of variable forcing terms in a more accurate way. Next, we consider extensions the simplified LBM algorithms for regimes of high values of relaxation times. In the final part, we introduce explicit immersed boundary algorithms for simulations of flows involving curved boundaries and whose accuracy is independent of the values of resistivity and viscosity coefficients.
	
	\section{Validations and benchmarks}\label{Section of results and discussion}

	In the next sections, we introduce some improvements in the simplified single-step algorithm (\ref{single-step algorithm complete}) and we show a series of numerical tests and validations for periodic flows in circular pipes with insulating boundaries in order to verify the suggested improvements. The numerical tests are described in more details as follows.
	
	 For examples involving stationary flows under the presence of a uniform magnetic field with insulating walls, as represented in Figure~\ref{Schematic representation pipe flow}, we compare the numerical solutions with the analytical solution derived by Richard R. Gold~\cite{gold1962magnetohydrodynamic} for a pipe flow submitted to a constant transverse magnetic field. The Gold's solutions for the streamwise  components of velocity and magnetic fields of the system (\ref{complete system of MHD}-\ref{incompressibility magnetic field}) are given by
\begin{eqnarray}
          U_x(r,\theta)&=& -\dfrac{R^2}{\nu Ha } \dfrac{\partial p}{\partial x} \left[  \cosh{(\alpha r \cos{\theta} )} \sum_{n=0}^{\infty} \varepsilon_n \dfrac{I^{'}_{2n}(\alpha)}{I_{2n}(\alpha)} I_{2n}(\alpha r)\cos{(2n\theta)} - \right. \nonumber \\
          &-& \left. \sinh{(\alpha r \cos{\theta})} \sum_{n=0}^{\infty} 2 \dfrac{I^{'}_{2n+1}(\alpha)}{I_{2n+1}(\alpha)} I_{2n+1}(\alpha r ) \cos{((2n+1)\theta)} \right], \label{Velocity Gold solution}\\
          B_x(r,\theta) &=& -\dfrac{1}{\sqrt{\eta \nu}}\dfrac{R^2}{2 Ha } \dfrac{\partial p}{\partial x} \left[    \sum_{n=-\infty}^{\infty} \left( \exp{(-\alpha r \cos{\theta})}\right. -\right. \nonumber\\ 
          &-& \left. \left. (-1)^n \exp{(\alpha r \cos{\theta})} \right) \dfrac{I^{'}_n(\alpha)}{I_n(\alpha)}I_n(\alpha r)\exp{(i n\theta)}- 2r\cos{\theta}  \right],\label{Magnetic field Gold solution}
\end{eqnarray}
where $\alpha=Ha/2$, $\epsilon_n$ equal 1 for $n=0$ and 2 for $n>0$. $I_n$ is the modified Bessel function of the first kind of order $n$ and $I'_n$ is the respective derivative. The Hartman number, in the context of the experiments of this article, is defined as $Ha=B_0R/\sqrt{\eta \nu}$, where $B_0$ is the characteristic magnetic field intensity and $R$ is the pipe radius. 
  
We also study the effects of non-stationary and transients flows by analysing the evolution of magnetic energy $E_m= \left\langle \frac{1}{2} |\*B|^2  \right\rangle$ and the kinetic energy $E_k=\left\langle \frac{1}{2} \rho |\*u|^2 \right\rangle$ (per unit of volume), where $\left\langle \cdot  \right\rangle$ denotes spatial averages within a cylinder with radius smaller than the radius of the pipe. The respective variations are given by~\cite{davidson2002introduction}
	\begin{eqnarray}\label{Energy balance equations}
	\dfrac{dE_m}{dt}&=& \eta \left\langle {\bf B} \cdot \nabla^2 {\bf B} \right\rangle - \left\langle  {\bf B} \cdot (\nabla \cdot ( {\bf u} {\bf B}- {\bf B} {\bf u} ))  \right\rangle   ,  \\[2pt]
	\dfrac{dE_k}{dt}&=&- \left\langle \*u  \cdot \nabla p \right\rangle +\mu \left\langle \*u \cdot \nabla^2 \*u \right\rangle   + \left\langle \*u \cdot ({\bf J} \times {\bf B} ) \right\rangle. \nonumber \label{Kinetic energy balance}
	\end{eqnarray}
The energy budget in \eqref{Energy balance equations} is analysed for constant and variable forcing term. For the study of unsteady forcing terms, we analyse	the effects of a variable pressure difference defined as follows
	\begin{equation}\label{variable force}
	\dfrac{\partial p}{\partial x}= F_0 \cos{\left(\dfrac{2\pi t}{T}\right)},
	\end{equation}
	where  $F_0$ is a reference force intensity and $T$ is the period.

	In order to be able to verify the Gold's solutions, we first need to introduce a set of improvements in the previous single-step algorithm given by~(\ref{single-step algorithm complete}) and~(\ref{Single step forcing term introdution}).  In the final part of the article, we also apply the suggested algorithms for examples involving non-uniform magnetic fields. All the numerical experiments will consider the so-called lattice Boltzmann units (lbu), a simple artificial set of units with grid spacing and time step verifying $\delta t=\delta x=1$.

  \section{Improved simplified single-step LBM algorithm }\label{Improved single step algorithm}

  \subsection{Improvement in the implementation of forcing terms}\hspace{0.2cm}

  For the proper consideration of the forcing terms in~(\ref{Boltzmann equation original}) in the simplified single-step algorithm~(\ref{Single step forcing term introdution}), we consider the introduction of a consistent forcing scheme that takes into consideration the discrete effects at the level of distribution functions, similar to the developments in~\cite{gao2021consistent,chen2018simplified}. In this section, we include the GZS forcing scheme~\cite{guo2002discrete} into the algorithm (\ref{single-step algorithm complete}). The BGK algorithm with the GZS scheme is expressed as
 \begin{equation}\label{BGK algorithm with GZS forcing}
     f_i({\bf x}+{\bf c}_i \delta t,t+\delta t)- f_i ({\bf x},t) = - \dfrac{f_i({\bf x},t)-f_i^{eq}({\bf x},t) }{\tau}+F_i \delta t
 \end{equation}
 where
 \begin{eqnarray}\label{FGS forcing term algorithm}
		F_i = \left(1-\dfrac{1}{2\tau}\right)w_i \left[ \dfrac{(\*c_i-\*u)}{c_s^2}+\dfrac{(\*c_i \cdot \*u )}{c_s^4} \*c_i \right] \cdot \*F_{ext},
\end{eqnarray}
with
\begin{eqnarray}
         \sum_i^N F_i &=& 0, \ \ \ \sum_i^N \*c_i F_i = \left( 1-\dfrac{1}{2\tau} \right) \*F_{ext},\\
         \sum_i^N \*c_i\*c_i F_i &=& \left( 1-\dfrac{1}{2\tau} \right) (\*u \*F_{ext}+\*F_{ext} \*u ).
\end{eqnarray}
  As pointed out by~\cite{gao2021consistent}, the application of the Chapman-Enskog expansion analysis in (\ref{BGK algorithm with GZS forcing}) gives rise to the following expression
  \begin{equation}\label{Pre-equation for descrete effects I}
      \sum_{i=1}^{N} \*c_i \left[ \dfrac{\partial f_i^{eq}}{\partial t} +\*c_i \cdot \nabla \left( f_i^{eq}+\left( 1-\dfrac{1}{2\tau} \right)  f_i^{(1)}+\dfrac{\delta t}{2} F_i \right) \right]=\*F_{ext},
  \end{equation}
  where this time
  \begin{equation}
      f^{(1)}\simeq - \tau \delta t D_{i} f_{i}^{eq}+\tau \delta t F_{i}.
  \end{equation}
  Follows that
  \begin{equation}\label{new non-equilibrium distribution forcing term}
      \sum_{i=1}^{N} \*c_i ( \*c_i \cdot \nabla f_i^{(1)}) \simeq - \sum_{i=1}^{N} \tau \delta t \*c_i (\*c_i \cdot \nabla ( \*c_i  \cdot \nabla f_i^{eq}))+ \tau \delta t \*c_i ( \*c_i \cdot \nabla F_i).
  \end{equation}
  Substituting (\ref{new non-equilibrium distribution forcing term}) into (\ref{Pre-equation for descrete effects I}), we obtain
  \begin{equation}\label{Pre-equation for descrete effects II}
      \sum_{i=1}^{N} \*c_i \left[ \dfrac{\partial f_i^{eq}}{\partial t} +\*c_i \cdot \nabla  \left(  f_i^{eq}-\left( \tau-\dfrac{1}{2} \right)\delta t  (\*c_i  \cdot \nabla f_i^{eq})  + \tau \delta t  F_i \right) \right]=\*F_{ext}.
  \end{equation}
  The extra term $ \*c_i (\*c_i \cdot \nabla)\tau \delta t F_i$ is associated with the lattice discrete effects that only appears for variable forcing terms. For this term, the following discretization based on isotropic finite differences~\cite{thampi2013isotropic} can be considered
  \begin{equation}\label{variable forcing terms}
      \sum_{i=1}^N \*c_i (\*c_i \cdot \nabla) \tau \delta t F_i \simeq  \sum_{i=1}^N  \dfrac{\tau }{2} \*c_i [ F_i(\*x+\*c_i \delta t,t)- F_i(\*x-\*c_i \delta t,t)].
  \end{equation}
  Therefore, the single-step algorithm for the velocity (\ref{Single step forcing term introdution}) should be rewritten as
		\begin{eqnarray}\label{Single-step algorithm with lattice discrete effect}
		\*u(\*x,t+\delta t)&= & \dfrac{1}{\rho(\*x,t+\delta t)} \sum_{i=1}^{N} \left\lbrace  \*c_i  f^{eq}_i(\*x-\*c_i \delta t,t)+\right. \\
		&+& \left. (\tau -1)[f^{eq}_i(\*x+\*c_i \delta t,t)-2f^{eq}_i(\*x,t)+ f^{eq}_i(\*x-\*c_i \delta t,t) ] \right\rbrace- \nonumber \\
		&-&  \dfrac{\tau \delta t}{2} \*c_i [ F_i(\*x+\*c_i \delta t,t)- F_i(\*x-\*c_i \delta t,t)] + \nonumber \\
		&+& \*F_{ext}(\*x,t) \delta t. \nonumber
		\end{eqnarray}
		
	With this improvement, it is possible to simulate more accurately multiple forms of external force interactions, including space- and time-dependent body forces, such as the Lorentz force~\cite{davidson2002introduction}
	\begin{equation}
	{\bf F}_{Lorentz}={\bf J} \times {\bf B}= (\nabla \times {\bf B} ) \times {\bf B},
	\end{equation}
	where the curl can be calculated by using isotropic finite differences~\cite{thampi2013isotropic}. Effects of magnetic fields can also be introduced by changing the equilibrium distribution~\cite{de2021one} in such a way that the divergence of the Maxwell stress tensor is implemented~\cite{de2019universal,de2021one}. This approach have not shown stable results in our numerical experiments for the case of non-uniform magnetic fields in simulations involving very small $R_m$. For this reason, in this article the forcing term approach is considered in all of the numerical experiments.

  \subsection{Boundary condition-enforced IBM}\hspace{0.2cm}
  
  In this section, in order to introduce the effects of curved boundaries in MHD flows, we consider the immersed boundary method (IBM). In this method a fixed Eulerian mesh is applied in which the flow field is resolved, while the immersed solid boundary is described by a set of discrete Lagrangian points distributed in the fluid domain. The flow variables resolved on the Eulerian mesh are corrected by a restoration force exerted from the solid boundary~\cite{kruger2017lattice}. In this article, we consider velocity and magnetic field corrections given by an extension of the boundary condition-enforced IBM based on the developments in~\cite{zhao2021efficient}, as we describe below.

	\begin{figure}[h!]
		\centering
	\subfigure[]{	\includegraphics[scale=0.62]{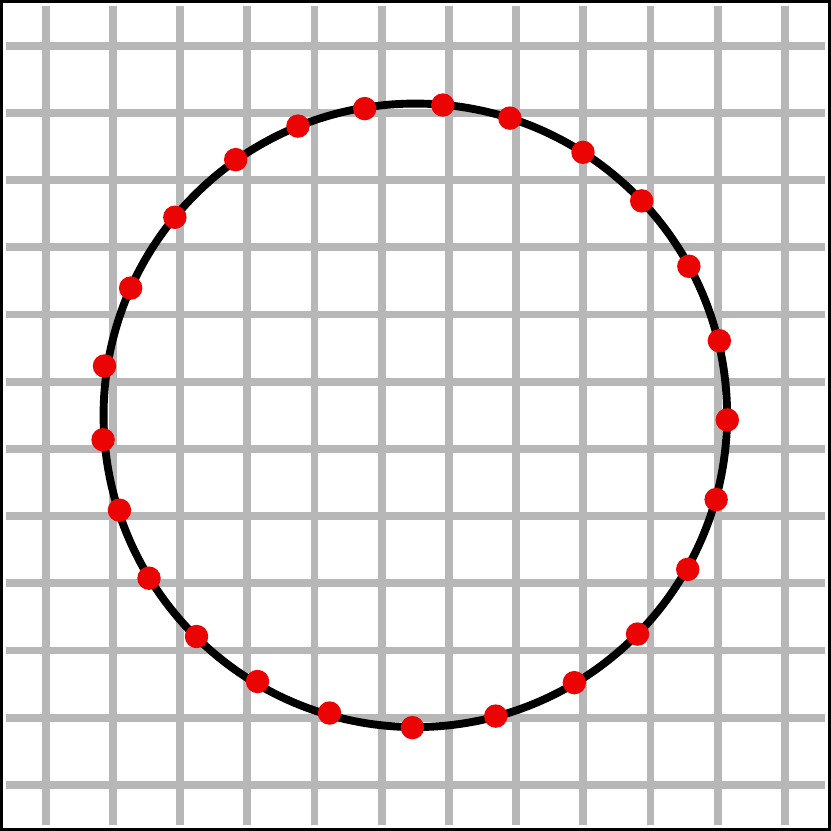}}\hspace{0.5cm}
	\subfigure[]{	\includegraphics[scale=0.67]{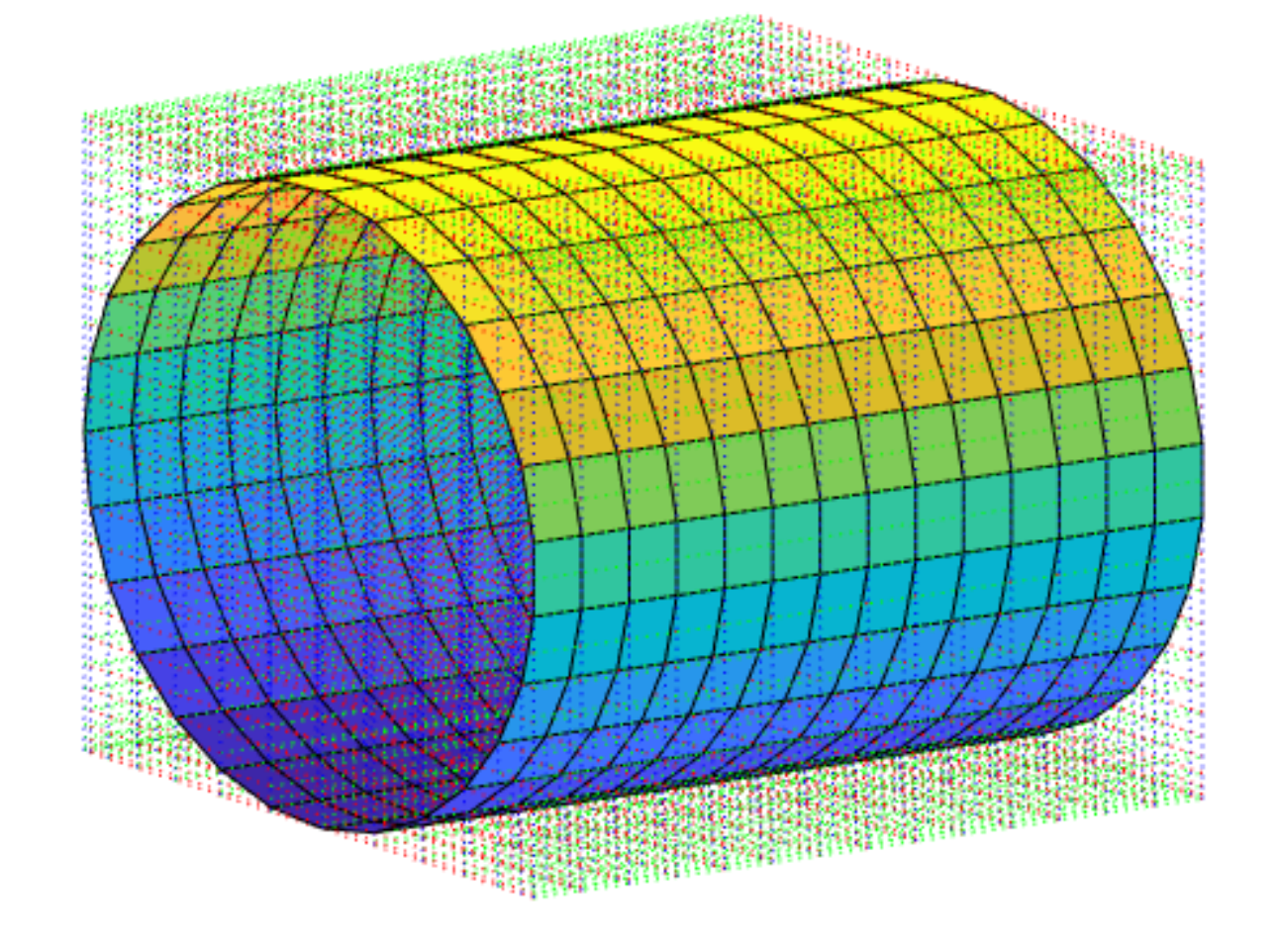}}
		\caption{(a) Cylinder with boundary markers (in red)
			positioned in the fluid domain. The Eulerian
			and the Lagrangian meshes are
			independent. (b) Schematic representation of the typical immersed boundary considered for the MHD pipe flows in this article.}
		\label{Schematic representation of Immersed Boundary article}	
	\end{figure}

  	In most of the IBM, the introduction of the effects of the boundaries is given by predictor-correction algorithm. In the predictor step, the LBM algorithm solves the following general system without boundary effects
	\begin{eqnarray}
	\dfrac{\partial \rho}{\partial t}+\nabla \cdot (\rho {\bf u} )&=&0\\
	\dfrac{\partial (\rho {\bf u} ) }{\partial t}+\nabla \cdot (\rho ({\bf u} \otimes {\bf u} ) )&=&-\nabla p +\nabla \cdot \left[ \mu (\nabla {\bf u}+\nabla {\bf u}^T )) \right] +{\bf F}_{ext}.
	\end{eqnarray}
	%where $\rho$, ${\bf u}$, $p$, and $\mu$ are the density, velocity, pressure and the dynamic viscosity of the fluid, respectively. ${\bf F}$ denotes the forcing term. 
	The effects of the boundaries are imposed as an  extra forcing term introduced in the corrector step as
	\begin{equation}\label{Corrector step IBM}
	\dfrac{\partial (\rho  {\bf u}) }{\partial t}= {\bf f},
	\end{equation}
	where $\*f$ is determined by the IBMs to reproduce the effects of the immersed objects. Since the forcing term $\*f$ is not considered in the prediction step, the intermediate velocity $\*u^*$ obtained in the predictor step must be corrected.
The corrector step (\ref{Corrector step IBM}) is discretized as
\begin{equation}
   \rho \delta {\bf u} = {\bf f} \delta t,
\end{equation}
where $\delta {\bf u}$ is the velocity correction. The corrected velocity is given by
\begin{equation}\label{Eulerian velocity corrections general}
    {\bf u} = {\bf u}^*+\delta {\bf u}.
\end{equation}

In order to calculate the corrections, interpolations  between the Lagrangian and the Eulerian meshes are usually made using the discrete
delta functions~\cite{kruger2017lattice}. In this article, we consider a different approach for the interpolation procedure, which was suggested by~\cite{amiri2020accuracy} in the context of 2D flows. In this work, the authors showed that the use of Lagrange polynomials, instead of numerical delta functions, gives significantly better results in terms of accuracy. More specifically, the velocity correction $\delta {\bf u}({\bf x}_i)$ at Eulerian mesh cell i is distributed from the velocity corrections $\delta {\bf U}( {\bf X}_j )$ at
Lagrangian points ${\bf X}_j$ by~\cite{amiri2020accuracy} using classical Lagrangian interpolation schemes given by
\begin{equation}\label{Interpolation of velocity corrections}
    \delta {\bf u}( {\bf x}_i )=\sum_{j=1}^{N} D({\bf x}_i - {\bf X}_j) \delta {\bf U}( {\bf X}_j ),
\end{equation}
and
\begin{equation}\label{Eulerian velocity as a function of Lagrangian velocity}
     {\bf u}( {\bf x}_i )=\sum_{j=1}^{N} D({\bf x}_i - {\bf X}_j)  {\bf U}( {\bf X}_j ),
\end{equation}
where $N$ is the total number of Lagrangian points and $D$ is accounts for a Lagrange velocity polynomial interpolation written as 
\begin{equation}\label{Lagrangian functions 1}
   D({\bf r})=D_x(r_x)D_y(r_y)D_z(r_z),
\end{equation}
where, for the purposes of this article, the coefficients are given by
\begin{equation}\label{Kernel functions for Lagrange interpolation I}
D_x(r_x)=
   \begin{cases}
   \dfrac{1}{2}(r_x+1)(r_x+2), \ \ \ \textrm{for} \ \ r_x \in [-3/2, \ -1/2],\\
   1-r_x^2,  \ \ \ \ \ \ \ \ \ \ \ \ \ \ \  \ \  \  \textrm{for} \ \  r_x \in [-1/2, \ 1/2],\\
   \dfrac{1}{2}(r_x-1)(r_x-2), \ \ \  \textrm{for} \ \ r_x \in [1/2, \ 3/2],
   \end{cases}
\end{equation}
and analogously for $D_y(r_y)$ and $D_z(r_z)$.
The use higher order Lagrange polynomials is possible~\cite{amiri2020accuracy}, but in the experiments of this article no significant differences were found by using them.

Analogously, the velocity $ {\bf  U}_b({\bf X}_j)$ at the Lagrangian point ${\bf X}_j$ can be interpolated from the corrected velocity $\*u$ at the Eulerian mesh points by using
\begin{equation}\label{Lagrangian velocity as a function of Eulerian velocities}
     {\bf U}_b( {\bf X}_j )=\sum_{i \in S(j)}^{} D({\bf x}_i - {\bf X}_j)  {\bf u}( {\bf x}_i ),
\end{equation}
where $S(j)$ is the set of neighboring Eulerian cells near the Lagrangian point ${\bf X}_j$ defined as
\begin{equation}\label{Set of Lagrangian points Sj}
   S(j)= \left\lbrace i,\left| \left| \dfrac{x_i-X_j}{h}  \right|, \left| \dfrac{y_i-Y_j}{h}  \right|, \left| \dfrac{z_i-Z_j}{h}  \right| \right. \leq 2 \right\rbrace,
\end{equation}
where $h$ is the grid spacing in the Eulerian mesh, which in this article is set to the unity  without loss of generality.  Substituting \eqref{Interpolation of velocity corrections} and \eqref{Eulerian velocity corrections general} into (\ref{Lagrangian velocity as a function of Eulerian velocities}), we obtain the following equation
\begin{equation}\label{Pre-equation for IBM system}
    {\bf U}_b ({\bf X}_j)= \sum_{i \in S(j)} D(  {\bf x}_i - {\bf X}_j) {\bf u}^* ( {\bf x}_i ) + \sum_{i \in S(j)} D(  {\bf x}_i - {\bf X}_j) \sum_{j=1}^{N}  D(  {\bf x}_i - {\bf X}_j) \delta {\bf U}( {X}_j ),
\end{equation}
where $\delta {\bf U}( {\bf X}_j )$ is an unknown velocity correction, ${\bf U}_b$ is an imposed velocity on the immersed boundary points and $\*u^*$ is known from the predictor step.  In a matrix form the relation (\ref{Pre-equation for IBM system}) is given by
\begin{equation}
     {\bf U}_b = {\bf D}\*u^* + {\bf D} {\bf D}^T \delta {\bf U},
\end{equation}
where

\begin{eqnarray}
{\bf U}_b&=&
\begin{bmatrix}
 {\bf U_b}({\bf X}_1)\\
\vdots\\
{\bf U_b}({\bf X}_N)
\end{bmatrix}, \ \ \ \  
{\bf u}^*=
\begin{bmatrix}
 {\bf u^*}({\bf x}_1)\\
\vdots\\
{\bf u^*}({\bf x}_M)
\end{bmatrix}, \ \ \ \ 
\delta {\bf U}=
\begin{bmatrix}
\delta {\bf U}({\bf X}_1)\\
\vdots\\
\delta {\bf U}({\bf X}_N)
\end{bmatrix}, 
\\
{\bf D}&=&
\begin{bmatrix}
D({\bf x}_1-{\bf X}_1) \hdots D({\bf x}_M-{\bf X}_1) \\
\vdots \ \ \ \ \ \ddots \ \ \ \ \  \vdots  \\
D({\bf x}_1-{\bf X}_N) \hdots D({\bf x}_M-{\bf X}_N)
\end{bmatrix}.
\end{eqnarray}
where $M$ is the total number of Eulerian points the sets $S(j)$, $j=1,\cdot \cdot \cdot,N$. The velocity correction $\delta {\bf U}$ is obtained by solving the system
\begin{equation}\label{General form a linear system velocity}
    {\bf A}\delta{\bf U}=\*b,
\end{equation}
where ${\bf A}={\bf D}{\bf D}^T \in \mathbb{R}^N \times \mathbb{R}^N$ and $\*b=\*U_b-\*D\*u^*$. The corresponding corrected velocity at the Eulerians nodes is given by
\begin{equation}\label{final IBM correction velocity}
    \*u = \*u^*+\delta \*u = \*u^*+\*D^T\delta \*U.
\end{equation}

It is important to mention that the matrices ${\bf D}$ and ${\bf D}^T$ are easily obtained but the inversion of a matrix ${\bf A}$ can be a non-trivial procedure. In the next, based on the developments in \cite{zhao2021efficient}, we discuss an explicit strategy to solve the problem (\ref{General form a linear system velocity}) which does not involve the direct inversion of the matrix ${\bf A}$.

\subsection{Explicit boundary condition-enforced IBM}\label{subsection about small curvature approximation}\hspace{0.2cm}

In a more explicit way, the system (\ref{General form a linear system velocity}) is given by
\begin{equation}\label{Reduced IMB formula}
    \sum_{ j } A_{i j} \delta {\bf U} ({\bf X}_j)={\bf b}_i,
\end{equation}
where
\begin{equation}
    A_{ij}=\sum_{k}D(\*x_k-\*X_i)D(\*x_k-\*X_j).
\end{equation}
Note that we only need to consider the non-zero values of the coefficients $A_{ij}$, i.e., in the summation in (\ref{Reduced IMB formula}) we only need consider  $j \in \{  A_{i j} \neq 0 \}$.
The momentum correction is then linearized in the vicinity of $\*X_i$  in the following form
\begin{equation}\label{linearized velocities IBM}
    \delta {\bf U}(\*X_j) = \delta {\bf U}(\*X_i) + \dfrac{\partial \delta {\bf U} }{\partial \*X }(\*X_i) d\*X_{ij}+O(\| d\*X_{ij} \|^2),
\end{equation}
where $d\*X_{ij}=\*X_j - \*X_i$. Assuming that the curvature of the immersed boundary is small in such a way it can be approximated by a straight wall in the vicinity of $\*X_i$~\cite{gsell2019explicit,zhao2021efficient}, it follows that
\begin{equation}\label{Lagrange polinomials and derivatives}
    \sum_{ j }^{} A_{ij} \dfrac{\partial \delta {\bf U} }{\partial \*X }(\*X_i) d\*X_{ij} \simeq 0,
\end{equation}
as a consequence of the properties of the interpolating function (\ref{Kernel functions for Lagrange interpolation I}). Substituting (\ref{linearized velocities IBM}) and (\ref{Lagrange polinomials and derivatives}) into (\ref{Reduced IMB formula}), we have
\begin{equation}\label{Reduced IMB formula II}
    \sum_{ j }^{} A_{i j} \delta {\bf U} ({\bf X}_i)={\bf b}_i,
\end{equation}
up to a second order error.
%It possible to see that the term $D(x_k-X_i)$ is non-zero if the Eulerian cell $k \in S(i)$, and $D(x_k-X_j)$ is non-zero if $k \in S(j)$.  
%When $k \in S(j)$ we have
%\begin{eqnarray}
 %          \|x_i - X_j\| = \|(2h,2h,2h)\| \simeq O(h)\\
  %         \|x_i - X_i\| = \|(2h,2h,2h)\| \simeq O(h)
%\end{eqnarray}
%end by using the triangular inequality, we have $\|d{\bf X}_{ij} \|=\| {\bf X}_i-{\bf X}_j\|<O(h)$. Under this hypothesis the following approximation holds
%\begin{equation}\label{Simplification for iBM}
%    \delta {\bf U}({\bf X}_j)=\delta {\bf U}({\bf X}_i)+O(h^2)
%\end{equation}
 Now note that the unknown correction $\delta {\bf U}({\bf X}_i)$ can now be moved out of the summation, which leads to the simplified system~\cite{zhao2021efficient}
\begin{equation}
    \delta  \*U({\bf X}_i) \sum_{j} A_{ij}= \*b_i, 
\end{equation}
or in a matrix form
\begin{equation}\label{Equation for Lagrangian velocity correction in explicit method}
    \begin{bmatrix}
\ddots  \ \ \  {\bf 0} \ \ \ \ \ {\bf 0} \\
{\bf 0} \ \ \ \ \  d_i \ \ \ \  {\bf 0}  \\
{\bf 0} \ \ \ \ \  {\bf 0}  \ \ \  \ddots
\end{bmatrix}  \delta {\bf U} = {\bf b}.
\end{equation}
where
\begin{equation}
    d_i=\sum_{  j \in  \{ A_{ij} \neq 0 \} } A_{ij}, \ \ i=1,\cdot \cdot \cdot,N.
\end{equation}
where $N$ is the number of the immersed boundary points. 
Substituting the solutions of (\ref{Equation for Lagrangian velocity correction in explicit method}) into (\ref{final IBM correction velocity}), follows that the corrected velocities in the Eulerian nodes is given by
\begin{equation}\label{Correction in the Eulerian nodes}
    {\bf u} = {\bf u}^* + {\bf D}^T \delta {\bf U} = {\bf u}^*+{\bf D}^T
    \begin{bmatrix}
\ddots  \ \ \   {\bf 0} \ \ \ \ \  {\bf 0} \\
{\bf 0} \ \ \ \ \  \frac{1}{d_i} \ \ \ \  {\bf 0}  \\
{\bf 0} \ \ \ \ \  {\bf 0}  \ \ \  \ddots
\end{bmatrix} 
{\bf b}.
\end{equation}
An interesting feature of this method is that it avoids the direct inversion of the matrix $\*A$ in  (\ref{General form a linear system velocity}), which can be computationally expensive, specially if moving boundaries are involved, which requires the inversion of A repeatedly.  The explicit character of (\ref{Correction in the Eulerian nodes}) also simplifies the implementation of the method on GPUs.

	\subsection{Explicit boundary condition-enforced IBM for magnetic field}\label{Explicit boundary IBM for magnetic field}\hspace{0.2cm}

In a analogous way, the introduction of the effects of Dirichlet boundary conditions are introduced in the equations for the magnetic field by considering a similar predictor-correction algorithm, where the intermediate flow variables obtained in the predictor step are then corrected by the IBMs in the subsequent corrector step. In the predictor step, we solve the system
\begin{eqnarray}\label{Predictor step magnetic field}
    	\dfrac{\partial \*B}{\partial t}+\nabla \cdot (\*u  \*B-\*B\*u)&=&\eta \nabla^2 \*B+{\bf Q},\\
     \nabla \cdot \*B &=& 0.
\end{eqnarray}
where ${\bf Q}$ denotes a general source term. In this method, the boundary effects are imposed as an extra source term introduced in following corrector step
\begin{equation}
    \dfrac{\partial \*B }{\partial t} = \*q,
\end{equation}
 where ${\bf q}$ is determined by the IBMs to include the effects of the magnetic fields generated by immersed objects. The corrector-step is discretized as 
 \begin{equation}
 \delta \*B = {\*q}\delta t,
 \end{equation}
 and the corresponding corrected magnetic field will be given by
 \begin{equation}
     \*B = \*B^*+\delta \*B,
 \end{equation}
 where $\*B^*$ is the magnetic field obtained in the predictor step (\ref{Predictor step magnetic field}). 
 Following the same steps as in the case involving the velocity field, it follows that
\begin{equation}\label{Correction in the Eulerian nodes magnetic field}
    {\bf B} = {\bf B}^*+{\bf D}^T
    \begin{bmatrix}
\ddots  \ \ \   {\bf 0} \ \ \ \ \  {\bf 0} \\
{\bf 0} \ \ \ \ \  \frac{1}{d_i} \ \ \ \  {\bf 0}  \\
{\bf 0} \ \ \ \ \  {\bf 0}  \ \ \  \ddots
\end{bmatrix} 
(\*B_b - \*D\*B^*),
\end{equation}
where $\*B_b$ is the imposed magnetic field on the immersed boundary points.
 
\begin{figure}[h!]
		\centering
		\includegraphics[scale=0.62]{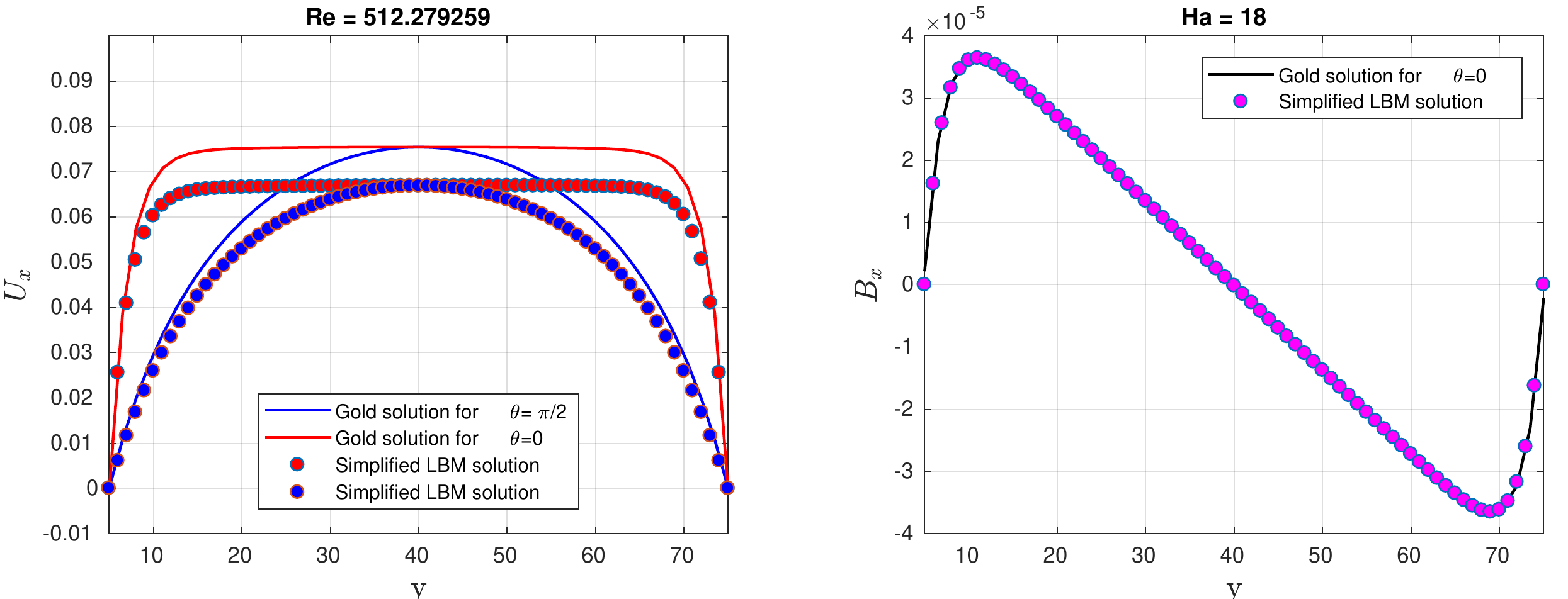}
		\caption{Comparison with the Gold's solutions by using the single-step LBM algorithms given by (\ref{single-step algorithm complete}) and (\ref{Single step algoritm for magnetic field}) with viscosity $\nu=0,004$, resistivity $\eta=10000$ and Hartman number $Ha=18$ in a pipe with radius $r=35$. The algorithm for the magnetic field is iterated $N_{mag}=12$ times before every update of the velocity field.  The resulting magnetic Prandtl number $Pr_m=4\times 10^{-7}$. In this simulation, we consider the immersed boundary corrections given by \eqref{Correction in the Eulerian nodes} and \eqref{Correction in the Eulerian nodes magnetic field}. It is possible to observe a significant mismatch between analytical and numerical solutions, which is corrected by the introduction of the new corrections \eqref{IBM correted for velocity field} and \eqref{IB corrected for magnetic field}, as shown in Figures~\ref{Different Hartmann circular pipes} and~\ref{First verification of the Quasi-static regime}.}
		\label{Problematic verification of the Quasi-static regime}	
	\end{figure}

 The use of the corrections (\ref{Correction in the Eulerian nodes}) and (\ref{Correction in the Eulerian nodes magnetic field}) gives accurate results if coupled with the single-step algorithm (\ref{single-step algorithm complete}) when the magnetic Prandtl number $Pr_m$ is closer to 1. For smaller values of $Pr_m$, some problems appears, as we can see in the Figure~\ref{IBM correction figures corrected}, where we performed simulations of two MHD flows using the algorithms (\ref{single-step algorithm complete}) and (\ref{Single step forcing term introdution}) with $Pr_m=0.1$ and $Ha=18$ in a computational grid with size $n_x \times n_y \times n_z=5\times80\times80$. The immersed boundary is approximated by a cylinder formed by small rectangular (almost squared) elements, as shown in Figure~\ref{Schematic representation of Immersed Boundary article}(b).  The number of elements is chosen in such a way that each element has an area close to $(\delta x)^2$, which is a common criterion for IB methods~\cite{kruger2017lattice}. It is possible to see a significant mismatch in the comparisons between the numerical solutions for $U_x$ and $B_x$ and the Gold's solutions (\ref{Velocity Gold solution}) and (\ref{Magnetic field Gold solution}). A similar mismatch also appears in the quasi-static regime as shown in Figure~\ref{Problematic verification of the Quasi-static regime}, where a simulation with $Pr_m=4\times 10^{-7}$ and $Ha=18$ with the same computational grid size is performed using some methods to be described in the next sections.  All this suggests that the accuracy of the corrections given by (\ref{Correction in the Eulerian nodes}) and(\ref{Correction in the Eulerian nodes magnetic field}) have some dependence with respect to the coefficients of viscosity and resistivity. It implies that for the simulations of the quasi-static approximation characterized by $Pr_m\ll1$, some improvements are needed. Strategies for the solution of this problem will be described in the next subsections.

\subsection{Viscosity-independent boundary condition-enforced IBM}\label{VIscosity independent IBM}\hspace{0.2cm}

In this section, we extend the previous results for IBM developed for the case where we present arbitrary magnetic Reynolds number.
In \cite{gsell2019explicit}, the authors suggested that the complete description of an immersed boundary problem also
involves the inclusion of non-dimensional IB force. More specifically,
in any physical configuration, the flow solution can be described by a set of non-dimensional physical quantities, as the non-dimensional pressure and velocity
\begin{equation}
    {\bf u}^*=\dfrac{{\bf u}}{U_r} \ \ \ p^*=\dfrac{p-p_r}{\rho_r U_r^2},
\end{equation}
where $U_r$, $p_r$ and $\rho_r$ are velocity, pressure and density of reference, respectively. 
In addition, a non-dimensional IB force is defined as
\begin{equation}\label{Immersed boundary force}
    \*f^*=\dfrac{{\bf f} D}{\rho_r U_r^2}.
\end{equation}

Consider two sets of dimensional quantities $(\rho_1,\*u_1,\*f_1)$ and $(\rho_2, \*u_2, \*f_2)$, which we call systems 1 and 2 respectively. Let us also consider that the reference densities and characteristic lengths are the same, i.e., $\rho_1=\rho_2=\rho_r$ (small Mach numbers assumption) and $L_1=L_2$. 
In this situation, if the both systems are solutions of the same physical problem, then the sets 1 and 2 results in the same set of non-dimensional quantities, which in our case implies in the same Reynolds, same Mach and same Froude numbers. In this case, denoting the reference velocities of the systems 1 and 2 by $U_{1}$ and $U_{2}$ respectively, it follows that the two systems are connected by the scaling factor defined as $\lambda=U_{2}/U_{1}$, which is also the viscosity ratio between configurations 1 and 2, i.e., $\lambda=\nu_2/\nu_1$. As a consequence, the following scaling laws are verified
\begin{equation}\label{Self similar properties}
    {\bf u}_1 = \dfrac{1}{\lambda} {\bf u}_2, \ \ \ {\bf f_1}=\dfrac{1}{\lambda^2} {\bf f}_2.
\end{equation}
The IB forces can be rewritten as
\begin{equation}
    {\bf f}_1=\rho_r \delta \*u_2 =\rho_r ({\bf u}_1 - {\bf u}_1^*), \ \ \ \  {\bf f}_2=\rho_r \delta \*u_2 =\rho_r ({\bf u}_2 - {\bf u}_2^*),
\end{equation}
 which leads to the the following equation
\begin{equation}\label{non self-similar IB}
    {\bf u}_1^* = \left( \dfrac{1}{\lambda} - \dfrac{1}{\lambda^2} \right) { \bf u}_2 +\dfrac{1}{\lambda^2} {\bf u}_2^*.
\end{equation}

Comparing (\ref{Self similar properties}) and (\ref{non self-similar IB}), we can observe that despite the fact that the physical quantities (\ref{Self similar properties}) exhibit self-similar scaling properties, the velocities corrected by the IBM cannot be directly rescaled using $\lambda$, because ${\bf u}_1^*$ has a dependence on ${\bf u}_2$. This property is one of the possible causes of the error shown in Figure~\ref{IBM correction figures corrected}. In the following, we describe the proper corrections that should be considered in order to introduce the correct IB adjustments. 

Let us denote the Lagrangian velocity corrections given by (\ref{Correction in the Eulerian nodes}) for the systems 1 and 2 as $\delta \*U_1$ and $\delta \*U_2$ respectively. The scaling verified in the Eulerian nodes should also be verified in the Lagrangian nodes, i.e.,  $\delta \*U_2=\lambda^2 \delta \*U_1$. Let us consider that the system 1 is a reference configuration that does not need scaling corrections. Using \eqref{General form a linear system velocity} it follows that
\begin{equation}\label{General form a linear system velocity corrected}
    \*D\*D^T \delta \*U_1 = {\bf U}_{b,1} - {\bf D}{\bf u^*_1}.
\end{equation}
As we already mentioned, the matrix $\*D\*D^T$ is usually ill-conditioned and its inversion is a non-trivial procedure, requiring some special techniques in order to approximate the inversion of $\*D\*D^T$. Let us consider, without loss of generality, the least square solution of \eqref{General form a linear system velocity corrected} written in terms of the pseudoinverse $(\*D\*D^T)^{\dag}$ with the representation formula given by 
\begin{equation}
    \delta {\bf U}_1 = (\*D\*D^T)^{\dag} ( {\bf U}_{b,1} - {\bf D}{\bf u^*_1}),
\end{equation}
and then
\begin{equation}\label{Scaling for Lagrangian corrections}
    \delta {\bf U}_2 = \lambda^2 (\*D\*D^T)^{\dag} ( {\bf U}_{b,1} - {\bf D}{\bf u^*_1}).
\end{equation}
Using (\ref{Self similar properties}) and (\ref{non self-similar IB}), it follows that from (\ref{Scaling for Lagrangian corrections}) we can derive
\begin{equation}
    \delta {\bf U}_2 = (\*D\*D^T)^{\dag} (\lambda {\bf U}_{b,2} + (1-\lambda)({\bf D} {\bf u}_2-{\bf D} {\bf u}^*_2)).
\end{equation}
Using (\ref{Correction in the Eulerian nodes}), we obtain
\begin{equation}\label{Original velocity correction u2}
    {\bf u}_2 = {\bf u}^*_2 + {\bf D}^T \delta {\bf U}_2,
\end{equation}
and finally, the IB force verifying the correct scaling properties will be given by
\begin{eqnarray}\label{IMB scaling for Lagrangian correction}
      \delta {\bf U}_2 &=& \lambda (\*D\*D^T)^{\dag}
 ({\bf U}_{b,2}- {\bf D} {\bf u}_2^* )  +  (1-\lambda) (\*D\*D^T)^{\dag}{\bf D} {\bf D}^T \delta {\bf U}_2. \nonumber \\
 &=& \lambda (\*D\*D^T)^{\dag}
 ({\bf U}_{b,2}- {\bf D} {\bf u}_2^* )  +  (1-\lambda) \*D \*D^{\dag} \delta {\bf U}_2,
\end{eqnarray}
where in the last equation we consider some general properties of pseudo-inverse matrices~\cite{golub2013matrix}. 
It is important to observe that the term $\*D \*D^{\dag}$ does not have to be the general identity matrix
$\*I$. Depending on the immersed boundary method,
 we may cancel the coefficient $\lambda$, but for some explicit velocity correction-based IBM, as the one described in this article, that is not the case. 
 
 Due to the properties of the interpolating functions~\eqref{Kernel functions for Lagrange interpolation I}, it follows that we can use power series and show that  one first approximation for $\*D^{\dag}$ is given by $\*D^T$ ~\cite{golub2013matrix,climent2001geometrical,tanabe1975neumann}.
  Using again \eqref{Equation for Lagrangian velocity correction in explicit method} and considering $\*D^{\dag} \simeq \*D^T$, we obtain
\begin{equation}
   {\bf D} {\bf D}^{\dag} \delta {\bf U}_2 \simeq {\bf D} {\bf D}^T \delta {\bf U}_2  \simeq \begin{bmatrix}
\ddots  \ \ \   {\bf 0} \ \ \ \  {\bf 0} \\
{\bf 0} \ \ \ \ \  d_i \ \ \ \  {\bf 0}  \\
{\bf 0} \ \ \ \  {\bf 0}  \ \ \  \ddots
\end{bmatrix} \delta {\bf  U}_2.
\end{equation}
Consequently, we can rewrite (\ref{IMB scaling for Lagrangian correction}) as
\begin{equation}\label{New corrections deltaU2 }
    \delta {\bf U}_2  = 
    \begin{bmatrix}
\ddots  \ \ \ \ \ \ \ \  {\bf 0} \ \ \ \ \ \ \ \ \ \ \ \ {\bf 0} \\
{\bf 0} \ \ \ \ \ \ \frac{\lambda}{1+d_i(\lambda-1)} \ \  \ \ \  \ {\bf 0}  \\
{\bf 0} \ \ \ \ \ \ \ \ \ \ \  {\bf 0} \ \ \ \  \ \ \ \ \ \ddots
\end{bmatrix} (\*D\*D^T)^{\dag}
({\bf U}_{b,2}- {\bf D} {\bf u}_2^*),
\end{equation}
where the term $(\*D\*D^T)^{\dag}({\bf U}_{b,2}- {\bf D} {\bf u}_2^*)$ in \eqref{New corrections deltaU2 } corresponds to the previous velocity correction obtained by finding the least-square solution of the system \eqref{General form a linear system velocity}. It is interesting to note that the form of the scalings in the matrix in the equation \eqref{New corrections deltaU2 } is very similar to the scalings obtained in \cite{gsell2019explicit} in the context of the direct forcing IBM, with the difference that in our work we found a matrix of scalings rather than a single scaling. 

Then, substituting (\ref{New corrections deltaU2 }) into (\ref{Original velocity correction u2}) and using \eqref{Equation for Lagrangian velocity correction in explicit method}, it follows that the new corrected velocity, considering the necessary scaling corrections, is be given by
\begin{equation}\label{IBM correted for velocity field}
    {\bf u}_2 = {\bf u}_2^* + {\bf D}^T 
      \begin{bmatrix}
\ddots  \ \ \ \ \ \ \ \ \ \  {\bf 0} \ \ \ \ \ \ \ \ \ \ \ \ \ {\bf 0} \\
{\bf 0} \ \ \ \ \ \ \frac{\lambda}{d_i(1+d_i(\lambda-1))} \ \ \ \ \ {\bf 0}  \\
{\bf 0} \ \ \ \ \ \ \ \ \ \ \ \ \ {\bf 0} \ \ \ \ \ \ \  \ \ \ \ddots
\end{bmatrix} {\bf b}.
\end{equation}
In the next subsection, we consider the introduction of similar corrections in the context of the explicit boundary condition-enforced IBM for the magnetic field equations. 

\subsection{Resistivity-independent boundary condition-enforced IBM}\hspace{0.2cm}

In this subsection, for the explicit IBM for the magnetic field described in the Subsection~\ref{Explicit boundary IBM for magnetic field}, we consider a procedure analogous to the case involving the velocity field. In this case, the two non-dimensional important physical parameters in this case are
\begin{equation}
    R_m=\dfrac{U_0L}{\eta}, \ \ \ \ Ha=\dfrac{B_0L}{\sqrt{\eta \nu}}.
\end{equation}
Consider two sets of dimensional quantities $({\bf u}_1,{\bf B}_1)$ and $({\bf u}_2, {\bf B}_2)$, which we also call systems 1 and 2 respectively. We also assume that the both sets are associated with the same physical system, which implies in the same set of non-dimensional quantities. The corresponding scaling factor will be given by $\lambda_{mag}=\eta_2/\eta_1$, which leads to the following relationships 
\begin{equation}
    {\bf B}_1 = \dfrac{1}{\lambda_{mag}} {\bf B}_2, \ \ \ {\bf q}_1 = \dfrac{1}{\lambda_{mag}^2} {\bf q}_2, 
\end{equation}
and similarly
\begin{equation}
    {\bf B}_1^* = \left(  \dfrac{1}{\lambda_{mag}}-\dfrac{1}{\lambda_{mag}^2}  \right) {\bf B}_2 + \dfrac{1}{\lambda_{mag}^2} {\bf B}_2^*,
\end{equation}
where $\*B_1^*$ and $\*B_2^*$ are magnetic fields obtained in the predictor step \eqref{Predictor step magnetic field}. The equation for the corrected magnetic field $\*B_2$ is then given by
\begin{equation}\label{IB corrected for magnetic field}
    {\bf B}_2 = {\bf B}_2^* + {\bf D}^T 
      \begin{bmatrix}
\ddots  \ \ \ \ \ \ \ \ \ \ \   {\bf 0} \ \ \ \ \ \ \ \ \ \ \ \ \   {\bf 0} \\
{\bf 0} \ \ \ \  \frac{\lambda_{mag}}{d_i(1+d_i(\lambda_{mag}-1))}  \ \ \ \  {\bf 0}  \\
{\bf 0}    \ \ \ \ \ \ \ \ \ \ \ \ \ {\bf 0} \ \ \ \ \ \ \   \ \ \ \ \ddots
\end{bmatrix} (\*B_{2,b} - \*D\*B^*_2),
\end{equation}
where $\*B_{2,b}$ is the imposed magnetic field on the immersed boundary points associated to the system configuration 2.

%The solutions of the system (\ref{General form a linear system}) gives the velocity correction $\delta {\bf u}({\bf x}_i)$ on the i-th Eulerian mesh cell, this corresponds to the force density
%\begin{equation}
%    {\bf f}_i = \dfrac{\rho \delta {\bf u}({\bf x}_i) }{\delta t}.
%\end{equation}
%The total force that the fluid receives from the immersed boundaries from the immersed boundary is ${\bf F}_{fluid}=\sum {\bf f}_i$, and the hydrodynamic force exerted on the immersed boundary is ${\bf F}_{wall}=-{\bf F}_{fluid}$~\cite{kruger2017lattice}.

\begin{figure}[h!]
		\centering
	\subfigure[]{	\includegraphics[scale=0.6]{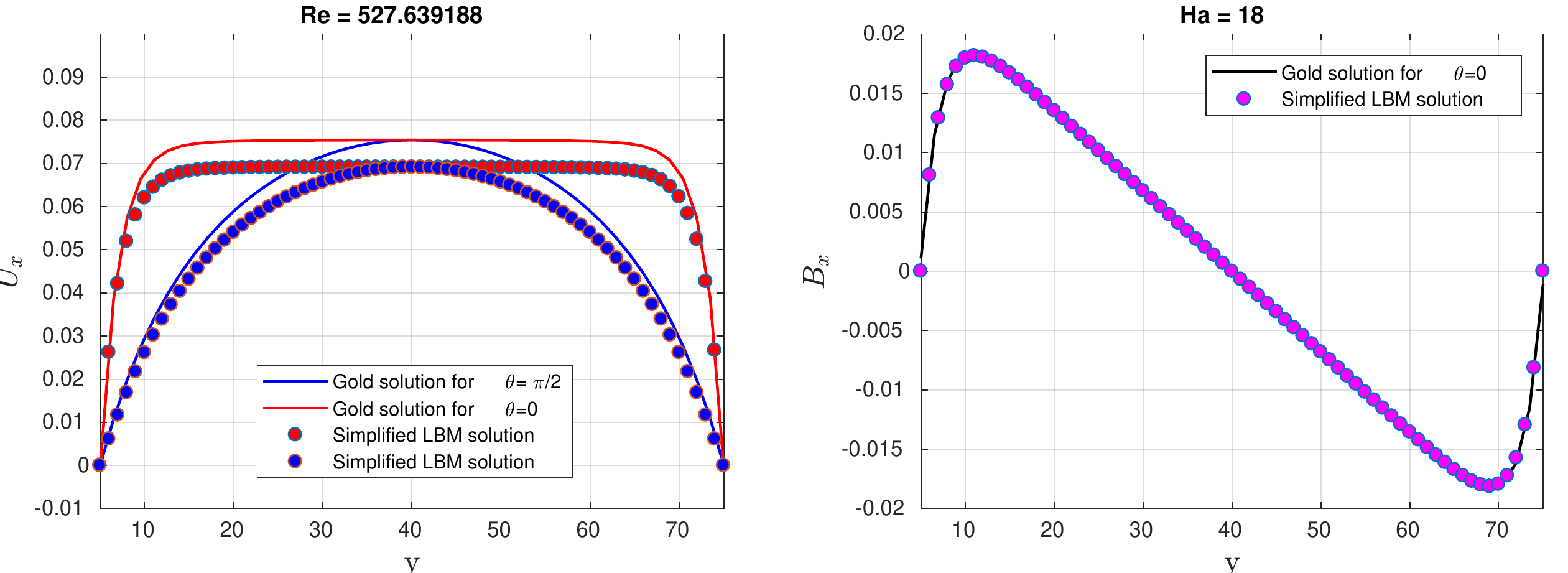}}
	\subfigure[]{
	\includegraphics[scale=0.6]{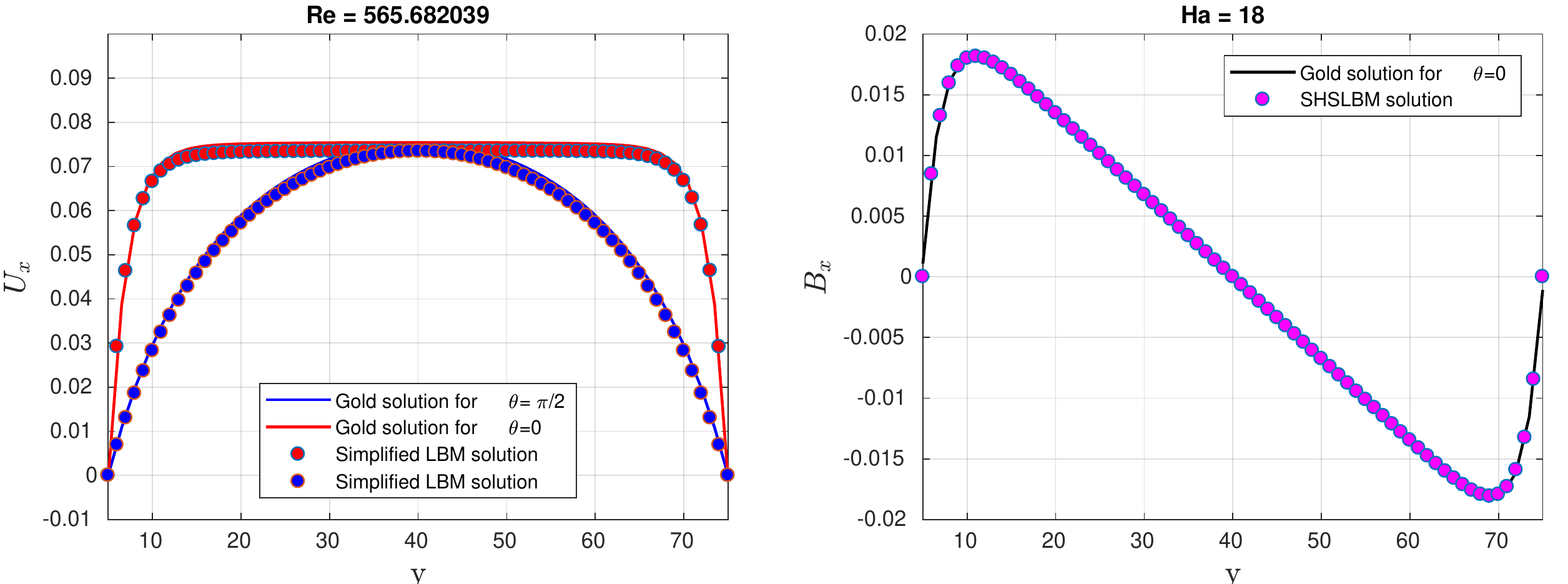}}
		\caption{Simulation of MHD pipe flows  under the presence of a transversal magnetic field with viscosity $\nu=0.004$, resistivity $\eta=0.04$ and $Ha=18$. In these experiments, we show the magnetic and velocity field profiles for a simulation with pipe radius $r=35$ and constant pressure difference $\frac{\partial p}{\partial x}=-3.7 \times 10^{-5}$, and we compare with the respective analytical solutions given by \eqref{Velocity Gold solution} and \eqref{Magnetic field Gold solution}. In (a) we consider the corrections given by \eqref{Correction in the Eulerian nodes} and \eqref{Correction in the Eulerian nodes magnetic field}. In (b) we consider the new corrections given by (\ref{IBM correted for velocity field}) and (\ref{IB corrected for magnetic field}) showing a much better agreement with the Gold's analytical solutions.}
		\label{IBM correction figures corrected}	
	\end{figure}

	An application of the new velocity and magnetic fields corrections by the formulas (\ref{IBM correted for velocity field}) and (\ref{IB corrected for magnetic field}) is shown in Figure~\ref{IBM correction figures corrected}, where we performed simulations of two MHD flows using the algorithms (\ref{single-step algorithm complete}) and (\ref{Single step forcing term introdution}) with $Pr_m=0.1$ and $Ha=18$. 
With the new corrections, we can observe a much better verification of the Gold's solutions (\ref{Velocity Gold solution}) and (\ref{Magnetic field Gold solution}).
	
	 \subsection{Stability improvements for high values of viscosity and resistivity}\hspace{0.2cm}

  In this section, we aim to extend range of stability of the previous simplified methods for regimes cassociated with high values of relaxation times.   
  The main idea is first to set the relaxation time $\tau=1$~\cite{zhou2020macroscopic,inamuro2002lattice} in the classical BGK algorithm (\ref{General Kinematic Boltzmann equation}) obtaining the so-called macroscopic lattice Boltzmann model given simply by
\begin{eqnarray}
\rho ({\bf x},t)&=& \sum_{\alpha} f_{\alpha}^{eq}({\bf x} - {\bf c}_i \delta t,t-\delta t ), \label{densities MAC}  \\
{\bf u}({\bf x},t)&=&\dfrac{1}{\rho ({\bf x},t)} \sum_{\alpha} {\bf c}_{\alpha} f_{\alpha}^{eq}({\bf x} - {\bf c}_i \delta t,t-\delta t ). \label{momentum Mac}
\end{eqnarray}
It is possible to show that the particle speed $c$ can be changed in such a way to include the effects of different viscosities $\nu$ as
\begin{equation}
c=\dfrac{6\nu}{\delta x},
\end{equation}
where $\delta t=\delta x/c$. In our applications, for the sake of simplicity, we always consider $\delta x=1$. Accordingly, the change in the particle speed $c$ also implies in the following changes in the lattice velocities of the D3Q27 scheme as
\begin{equation}
    \*c_{x}=(0, -c, 0, 0, -c, -c, -c, -c, 0, 0, -c, -c, -c, -c, c, 0, 0, c, c, c, c, 0, 0, c, c, c, c)^T,
\end{equation}
\begin{equation}
    \*c_{y}=(0, 0, -c, 0, -c, c, 0, 0, -c, -c, -c, -c, c, c, 0, c, 0, c, -c, 0, 0, c, c, c, c, -c, -c)^T,
\end{equation}
\begin{equation}
    \*c_{z}=(0, 0, 0, -c, 0, 0, -c, c, -c, c, -c, c, -c, c, 0, 0, c, 0, 0, c, -c, c, -c, c, -c, c, -c)^T.
\end{equation}
The algorithm formed by (\ref{densities MAC}) and (\ref{momentum Mac}) is particularly efficient and stable for flow simulations with small and moderate Reynolds numbers. High Reynolds numbers usually will require a very small $\delta x$, which implies in a substantial increase of the number of points in the computational grid. In this article, this algorithm is suggested as an extension for $\tau \geq 1$ of the single-step algorithm (\ref{single-step algorithm complete}). Actually, it can be considered an extension for any other simplified method that also have problems for high values of relaxation times.

In this article, we also extend the idea of the macroscopic LBM algorithm for the magnetic field equations \eqref{ADE for magnetic field} and \eqref{incompressibility magnetic field}. 	Substituting  $\tau_m=1$ in (\ref{original LBM for magnetic field}), we obtain the following algorithm
	\begin{eqnarray}\label{Single step algoritm for magnetic field}
	B_x({\bf x},t)&=&\sum_{i} g_{xi}^{eq}({\bf x}-{\bf c}_{i} \delta t ,t-\delta t), \nonumber \\
	B_y({\bf x},t)&=&\sum_{i} g_{yi}^{eq}({\bf x}-{\bf c}_{i} \delta t ,t-\delta t),\\
	B_z({\bf x},t)&=&\sum_{i} g_{zi}^{eq}({\bf x}-{\bf c}_{i} \delta t ,t-\delta t). \nonumber
	\end{eqnarray}
Recall the formula for the resistivity $\eta$ as a function of the relaxation time $\tau_m$ given by
	\begin{equation}\label{Resistivity as a function of the relaxation time}
	\eta = c_s^2\left(\tau_m -\dfrac{1}{2} \right) \delta t =  \dfrac{c^2}{3}\left(\tau_m -\dfrac{1}{2} \right) \delta t.
	\end{equation}
Introducing $\tau_m=1$ in \eqref{Resistivity as a function of the relaxation time}, we obtain
	\begin{equation}\label{Relationship between resistivity and time step}
\eta = \dfrac{c^2}{6} \delta t =\dfrac{c}{6} \delta x=\dfrac{(\delta x)^2}{6 \delta t},
	\end{equation}
and considering $\delta x=1$, we have $\eta = c/6$.

\begin{figure}[h!]
		\centering
	\subfigure[]{\includegraphics[scale=0.5]{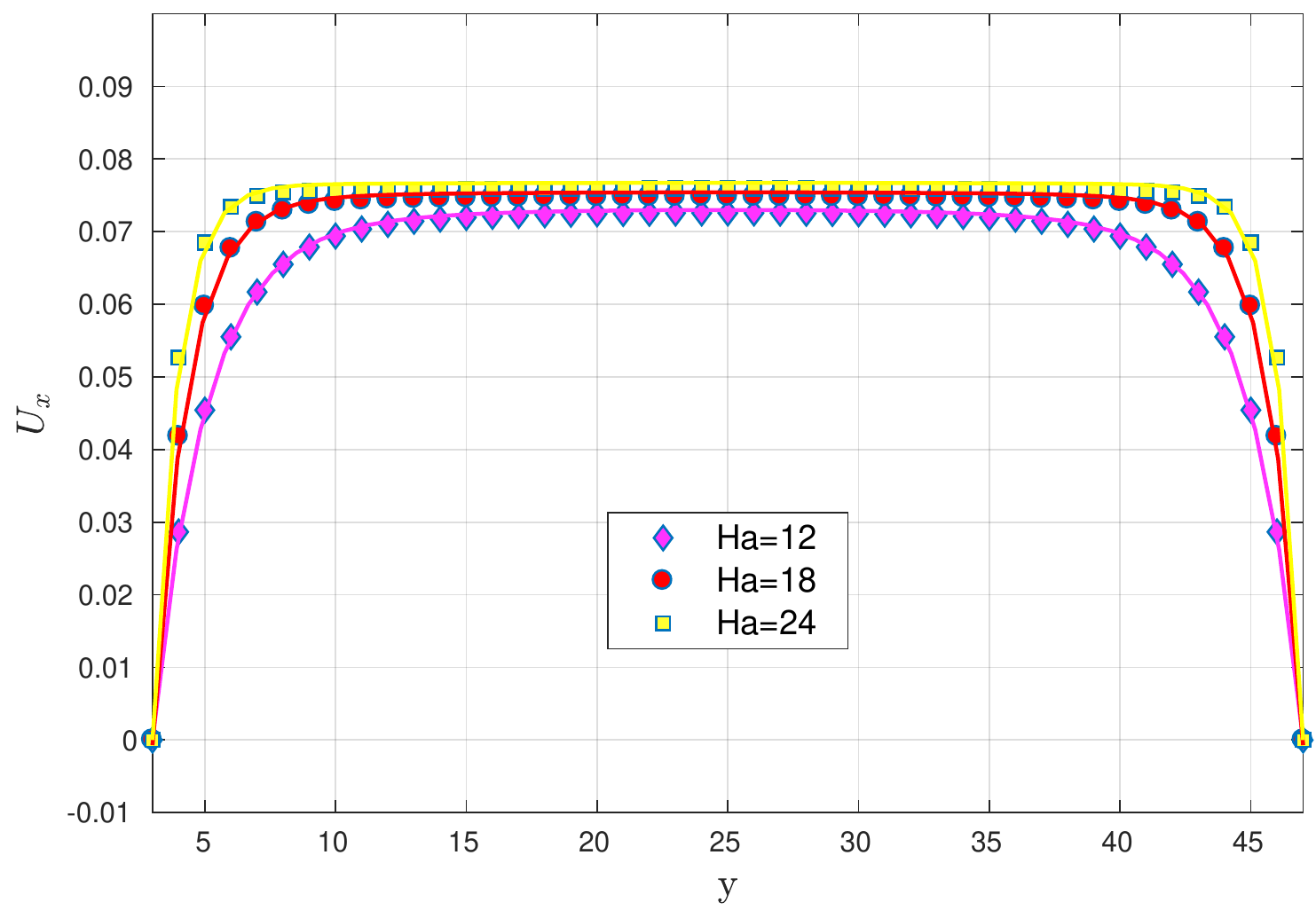}}\hspace{1cm}
	\subfigure[]{	\includegraphics[scale=0.5]{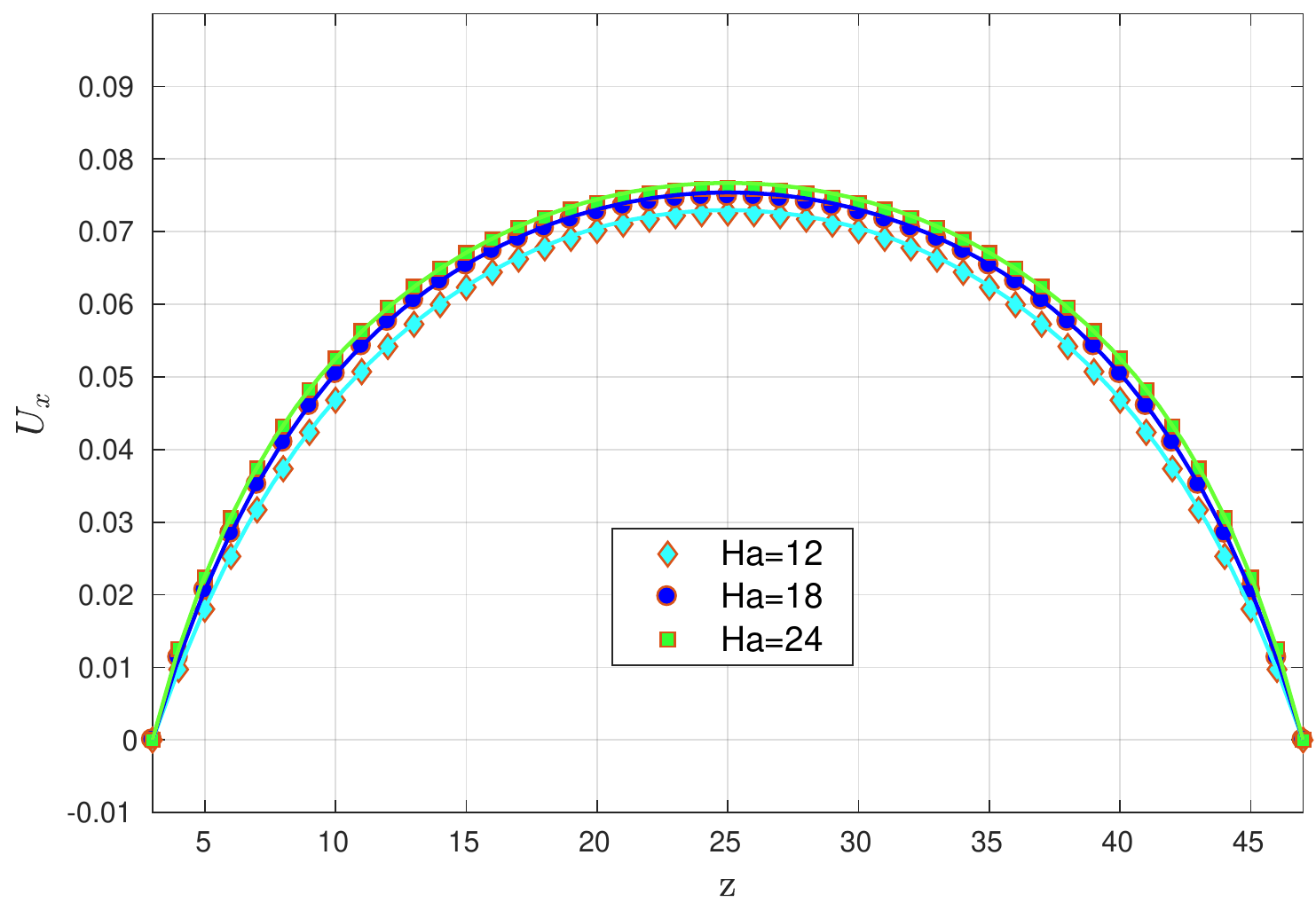}}\\
	\subfigure[]{	\includegraphics[scale=0.5]{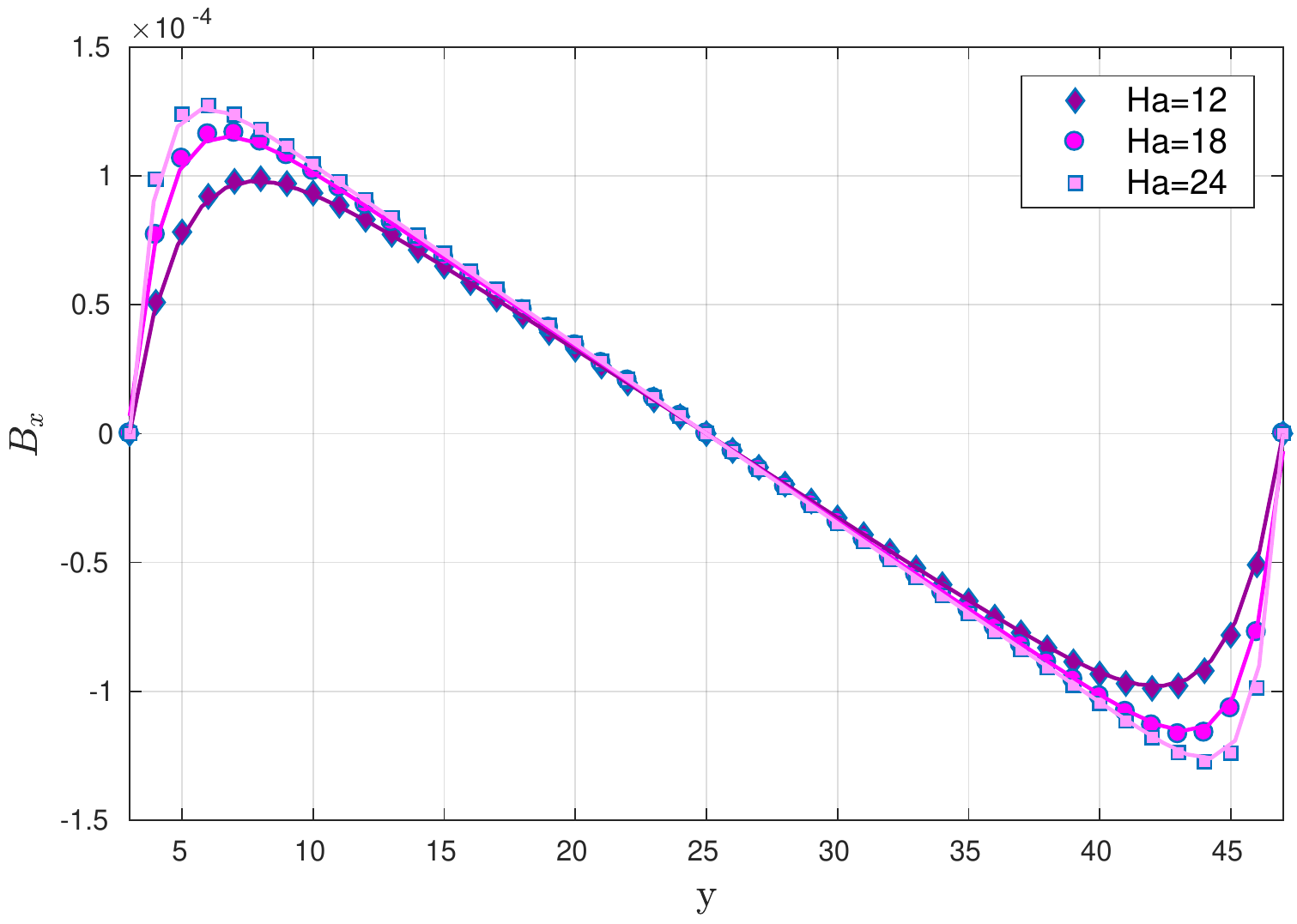}}
		\caption{Simulation of a MHD pipe flow submitted to a constant transverse magnetic field  with a pipe radius $r=22$, resistivity $\eta=10000$ and viscosity $\nu=0.04$. The results for three different values of Hartman $Ha=12,18$ and $24$ are shown. In picture (a), the velocity profiles corresponding to the Gold's solution for $\theta=\pi/2$; in picture (b), the solutions corresponding to $\theta=0$, and in (c), the magnetic field profiles for $\theta=\pi/2$. The continuous lines are the respective analytical solutions for each value of $Ha$.}
		\label{Different Hartmann circular pipes}	
	\end{figure}

\begin{figure}[h!]
		\centering
		\subfigure[]{
		\includegraphics[scale=0.6]{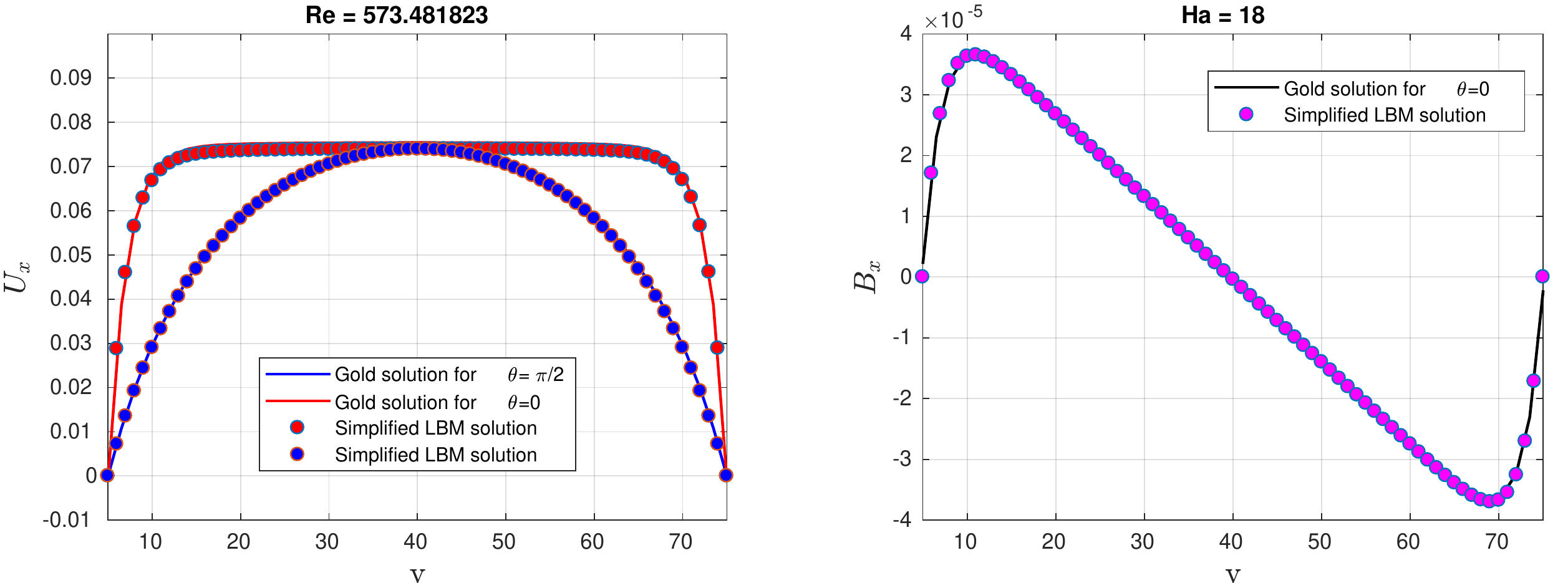}}\hspace{0.5cm}
		\subfigure[]{
		\includegraphics[scale=0.6]{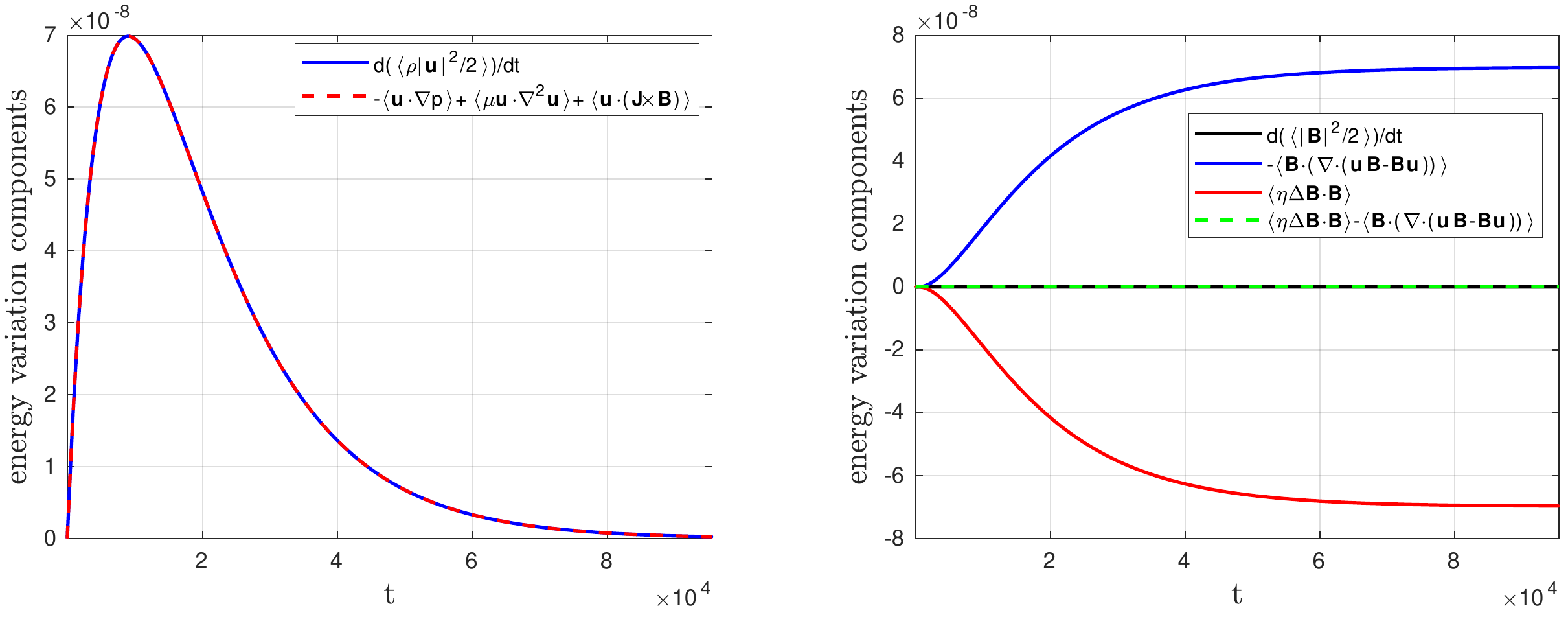}}
		\caption{Verification of the Gold's solutions by using the single-step LBM algorithms given by (\ref{single-step algorithm complete}), \eqref{Single step forcing term introdution} and (\ref{Single step algoritm for magnetic field}), with viscosity $\nu=0,004$, resistivity $\eta=10000$ and Hartman number $Ha=18$. The simulation is performed in a pipe with radius $r=35$.  The resulting magnetic Prandtl number equal to $Pr_m=4\times 10^{-7}$. In (a) we show the velocity and magnetic field profiles verifying the Gold's solutions (\ref{Velocity Gold solution}) and (\ref{Magnetic field Gold solution}), and in (b), we show the verification of the energy balance given by the equations (\ref{Energy balance equations}).}
		\label{First verification of the Quasi-static regime}	
	\end{figure}

	The algorithm given by~(\ref{Single step algoritm for magnetic field}) solves (\ref{ADE for magnetic field}) and (\ref{incompressibility magnetic field}) for a wide range of $\eta$ values, but similarly to the algorithm given by  (\ref{densities MAC}) and (\ref{momentum Mac}) for the velocity field, this algorithm is not practical for small values of resistivity, but is very suitable for the values of resistivity associated with the quasi-static approximation (\ref{quasi-static approximation}). 
	The idea in this article is to set the $\delta x=1$ in (\ref{Relationship between resistivity and time step}) (velocity and magnetic fields are solved in the same computational grid) and obtain $\delta t=1/6\eta$. It implies that if $\eta>1/6$, then the algorithm~(\ref{Single step algoritm for magnetic field}) should be iterated a few times before every update of the single-step algorithm given by~(\ref{single-step algorithm complete}) and (\ref{Single step forcing term introdution}) for the momentum equation. The number of iterations $N_{mag}$ for the algorithm (\ref{Single step algoritm for magnetic field}) can be defined as
\begin{equation}\label{Number of iterations magnetic field}
    N_{mag}=\ceil*{\frac{1}{ \delta t}},
\end{equation}
where the function $\ceil*{\cdot}$ denotes the smallest integer number greater or equal to $1/\delta t$. In many applications, very high values of resistivity generate a prohibitive value of $N_{mag}$, but in this situations we can work with some kind of effective number of iterations, as we shown in details in the Section~\ref{Strategies for very high values of resistivity}. The result of this strategies is shown in Figures~\ref{Different Hartmann circular pipes} and \ref{First verification of the Quasi-static regime}.

For a significant high values of $\eta$, the algorithm given by (\ref{Single step algoritm for magnetic field}) converges to the equation
	\begin{equation}
	     \eta\nabla^2 {\bf B} =  \nabla \cdot (\*u \*B-\*B\*u).
	\end{equation}
as a natural asymptotic limit.  
A verification of the proposed single-step algorithm is shown in Figure~\ref{Different Hartmann circular pipes}, where a pipe flow submitted to a uniform transverse magnetic field is implemented for three different values of Hartmann numbers, $Ha=12, 18$ and $24$, with $\eta=10000$ and $\nu=0,04$. Periodic boundary conditions are considered in the streamwise directions with Dirichlet boundary conditions at the walls of the pipe. All the simulations are initialized from zero velocity. The numerical solutions are compared with the Gold's solutions given by~(\ref{Velocity Gold solution}) and~(\ref{Magnetic field Gold solution}) for the values of $\theta=0$ and $\theta=\pi/2$, showing a good agreement. In Figure~\ref{First verification of the Quasi-static regime}, we analyze in more details the simulation with $Ha=18$, showing that the solution not only verifies the Gold's solution, but also the initial transient regime accurately verifies the energy balance given by~(\ref{Energy balance equations}).

For most of the experiments in this article, we set $Ha=18$, which we consider a representative value for the simulations with Hartman numbers between 1 and 30, in the sense that no significant differences have been verified by changing the values of $Ha$ in this range.  For higher values of Hartman numbers, grid refinements, especially close to the boundaries may be needed to handle the intensification of the Hartman layers~\cite{pattison2008progress}, for example.

	In the references~\cite{pattison2008progress,premnath2009steady}, the authors consider the introduction of extra parameters $\chi$ and $\gamma$ and use the traditional BGK algorithm (\ref{original LBM for magnetic field}) to solve the following equation 
 	\begin{equation}\label{preconditioning previous results}
    \dfrac{\partial \*B}{\partial t}+\dfrac{\chi}{\gamma}\nabla \cdot (\*u\*B-\*B\*u)=\dfrac{\eta}{\gamma} \nabla^2 \*B,
   \end{equation}
   which has a stationary solution given by
   \begin{equation}\label{preconditioning previous results QS}
   \nabla \cdot (\*u\*B-\*B\*u)=\dfrac{\eta}{\chi} \nabla^2 \*B.
   \end{equation}
The parameter $\chi$ can be set to archive the desired magnetic Prandtl number $Pr_m$ and the parameter $\gamma$, usually much smaller then 1, helps to increases the convergence rate to steady state solutions. The same strategy can also be applied for the single-step algorithm~(\ref{single-step algorithm complete}) as well. Originally in~\cite{pattison2008progress}, this procedure is mostly considered for steady states solutions, but its applicability for general flow regimes with $R_m\ll1$ is not clear. In this article, the algorithm~(\ref{Single step algoritm for magnetic field}) gives a more direct route towards simulations with very small $R_m$ with a much simpler and stable algorithm, and without the need of the introduction of extra parameters.

	\begin{figure}[h!]
		\centering
	\subfigure[]{\includegraphics[scale=0.6]{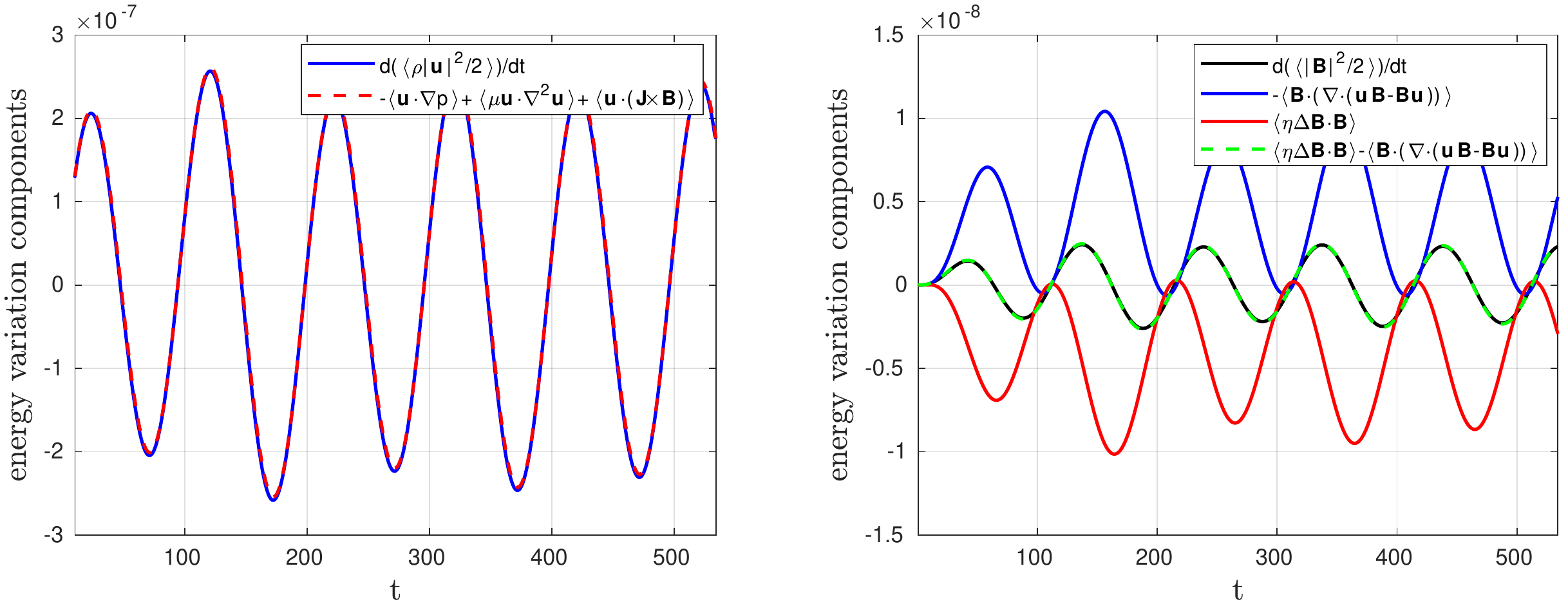}}
	\subfigure[]{	\includegraphics[scale=0.6]{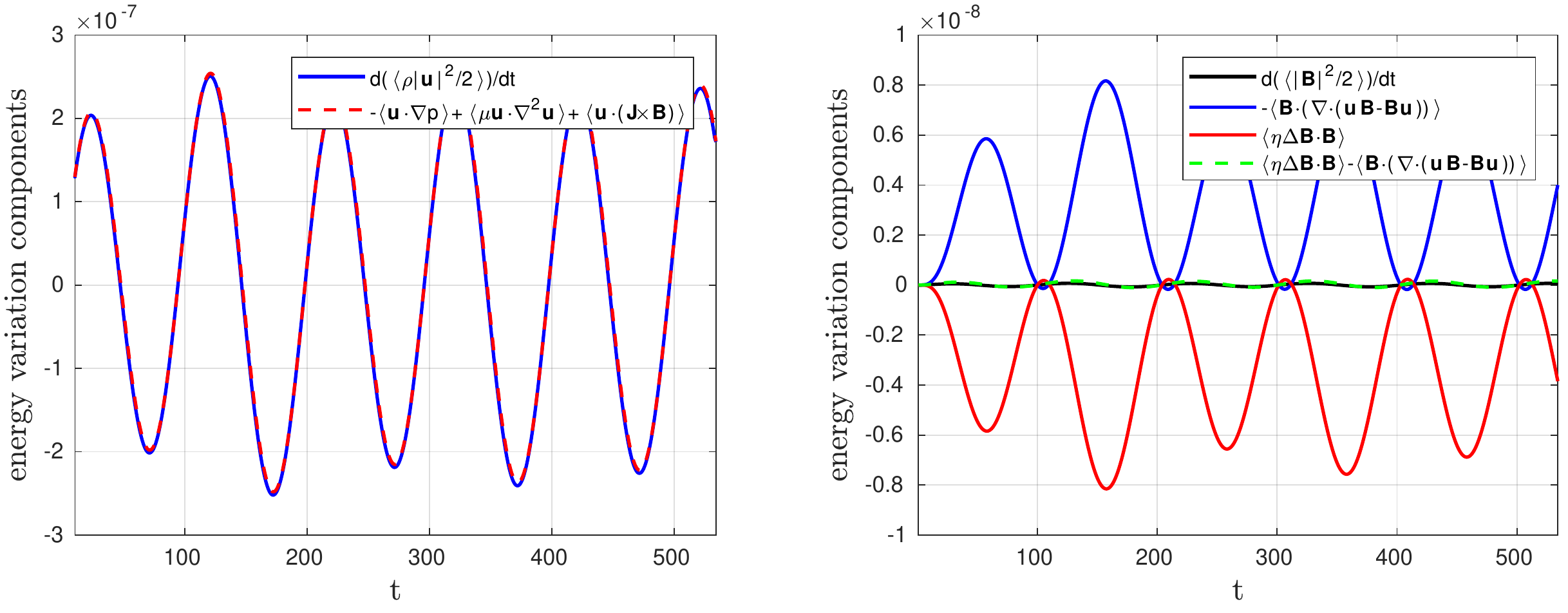}}
		\caption{In this figure we show the transition to the quasi-static regime. In (a) we show a simulation of a MHD flow in a circular pipe with radius $r=22$, $Ha=18$, $\eta=2$ and $\nu=0.08$ submitted to $\partial p/\partial x=-2.4\times10^{-4}\cos(2\pi t/200)$ with $\ceil*{\frac{1}{ \delta t}}=N=12$ iterations for (\ref{Single step algoritm for magnetic field}). In (b), we set $\eta=45$, which implies in $\ceil*{\frac{1}{ \delta t}}=N=270$ iterations for the algorithm (\ref{Single step algoritm for magnetic field}). We can observe in (b) the complete damping of the time derivative of the magnetic field, which is characteristic of the quasi-static approximation.}
		\label{Transition to the quasi-static regime}	
	\end{figure} 
	
	\subsection{Strategies for very high values of  resistivity}\label{Strategies for very high values of resistivity}\hspace{0.2cm}
	
	The formula~(\ref{Number of iterations magnetic field}) gives the necessary number of iterations for the convergence of the algorithm~(\ref{Single step algoritm for magnetic field}). Naturally, if the value of the resistivity is too high, the number of iterations becomes prohibitive for numerical purposes. In this subsection, we shown some strategies for the solution of this problems.
	
\begin{figure}[h!]
		\centering
			\subfigure[]{
		\includegraphics[scale=0.55]{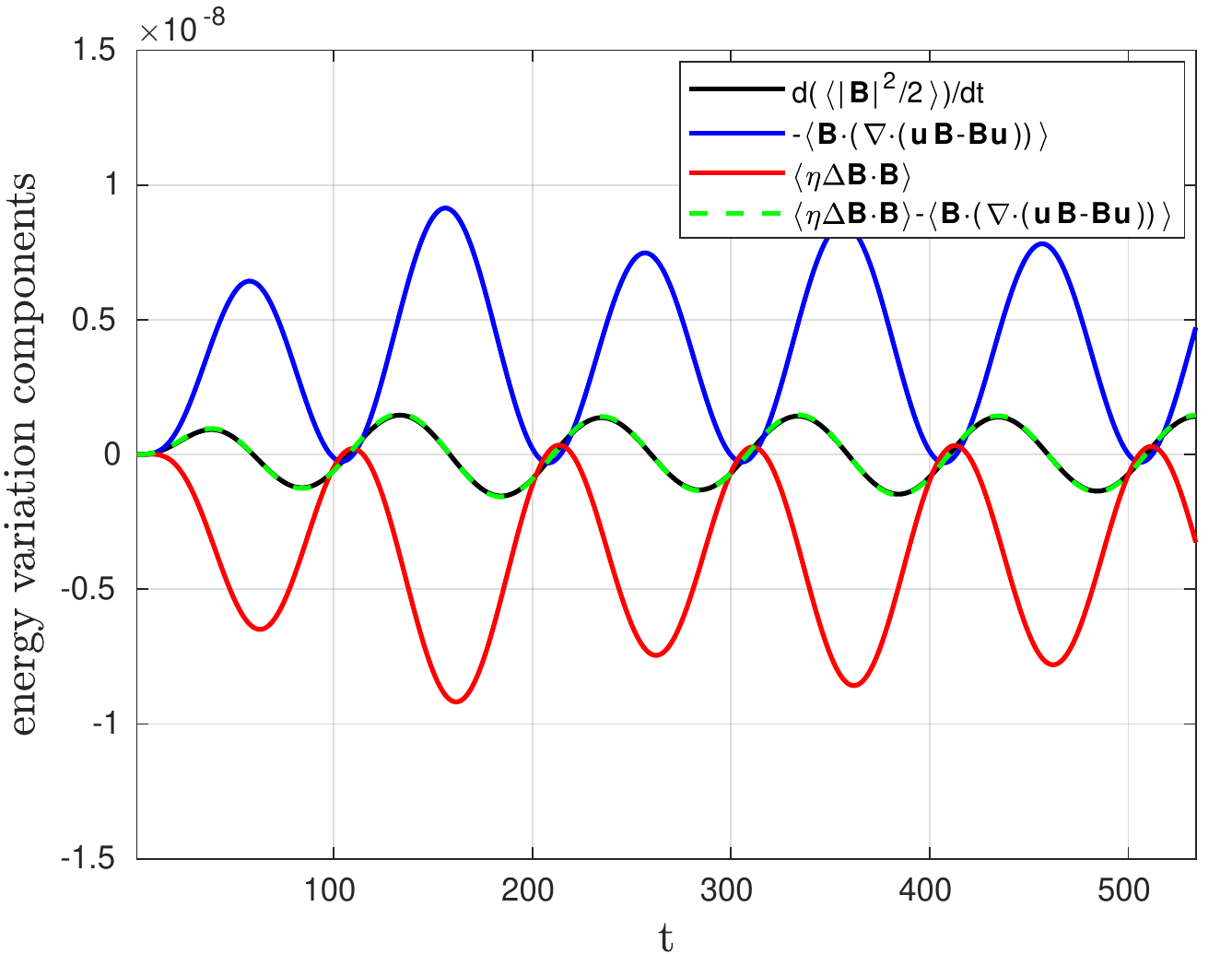}}\hspace{0.5cm}
			\subfigure[]{
		\includegraphics[scale=0.55]{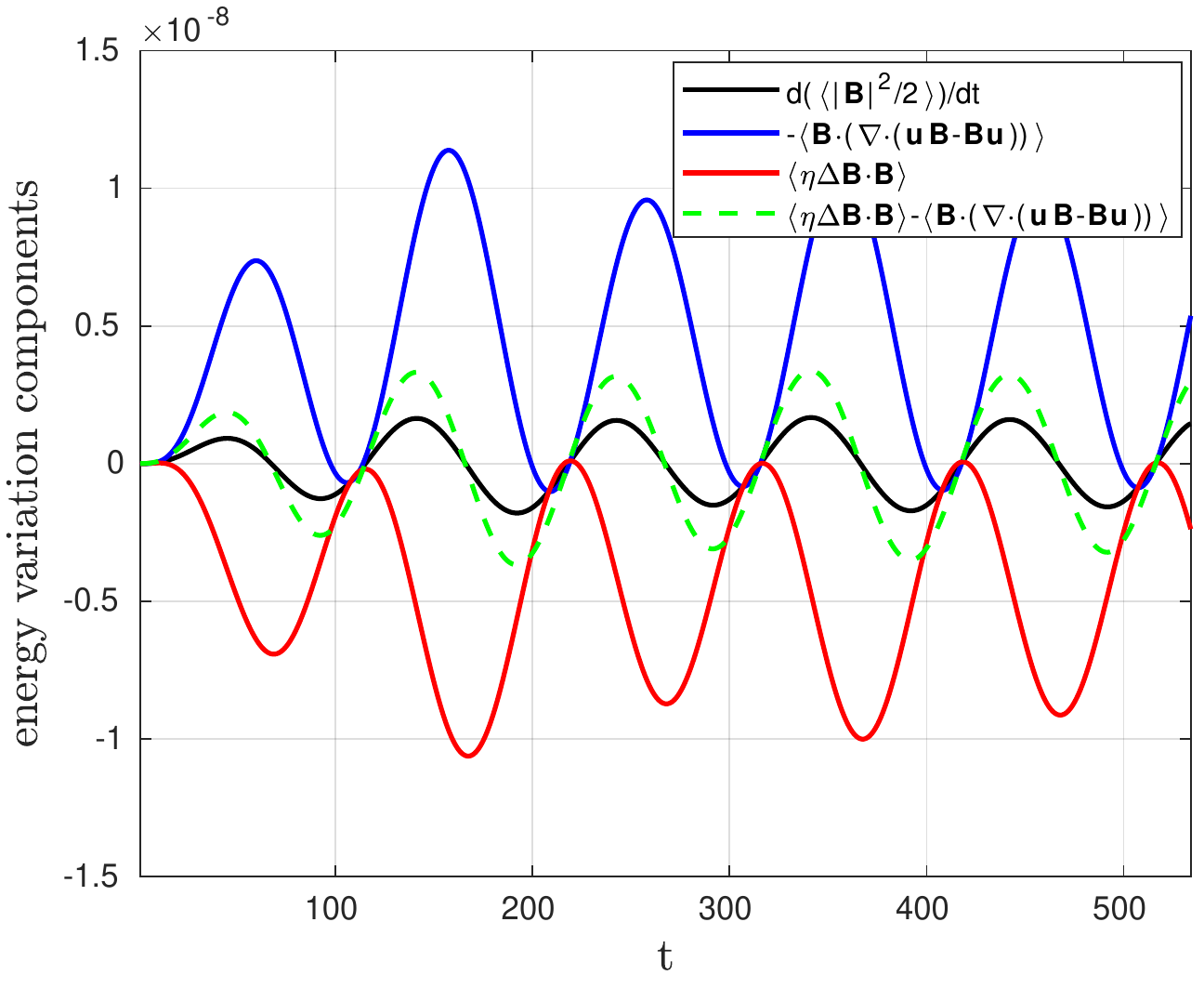}}\hspace{1cm}
		\caption{Simulation with $\eta=3$, $\nu=0.08$, $Ha=18$ and pipe radius $r=22$. The variable pressure difference in this case is given by and $\frac{\partial p}{\partial x}=-1.9\times10^{-4}\cos{(2\pi t/200)}$. In picture (a), a simulation using the algorithm (\ref{magnetic field quasi-static acelerated}) is shown, where the convergence is reached with $N=9$ iterations, which is half of the number of iterations recommend by the formula (\ref{Number of iterations magnetic field}) for the algorithm (\ref{Single step algoritm for magnetic field}), whose result with the same number of iterations is shown in picture (b).}
		\label{Acelerating convergence}	
	\end{figure}

We first consider a small modification in the equilibrium distributions given by (\ref{Equilibrium distribution for magnetic field I}), (\ref{Equilibrium distribution for magnetic field II}) and (\ref{Equilibrium distribution for magnetic field III}) as follows.  For ${\bf c}_1=[0, \ 0, \ 0]$, let us introduce an extra coefficient $\alpha$ as 
\begin{eqnarray}
g_{1x}^{eq} = (1-w_1)B_x+\alpha B_x, \nonumber \\
g_{1x}^{eq} = (1-w_1)B_y+\alpha B_y,\\
g_{1x}^{eq} = (1-w_1)B_z+\alpha B_z. \nonumber
\end{eqnarray}
and for the other velocities $i=2,\cdot \cdot \cdot,N$, consider
\begin{eqnarray}
g_{ix}^{eq} = w_i\left[B_x+\dfrac{c_{iy}}{c_s^2} (u_y B_x - u_x B_y)+ \dfrac{c_{iz}}{c_s^2} ( u_z B_x - u_x B_z) )   \right], \nonumber \\
g_{ix}^{eq} = w_i\left[By+\dfrac{c_{ix}}{c_s^2} (u_x B_y - u_y B_x)+ \dfrac{c_{iz}}{c_s^2} ( u_z B_y - u_y B_z) )   \right],\\
g_{ix}^{eq} = w_i\left[Bz+\dfrac{c_{ix}}{c_s^2} (u_x B_z - u_z B_x)+ \dfrac{c_{iy}}{c_s^2} ( u_y B_z - u_z B_y) )   \right], \nonumber
\end{eqnarray}
with the following small modification in the algorithm (\ref{Single step algoritm for magnetic field}) given by
\begin{eqnarray}\label{magnetic field quasi-static acelerated}
  B_x({\bf x},t)&=&\dfrac{1}{\alpha}\sum_{\alpha=1}^9 g_{x\alpha}^{eq}({\bf x}-{\bf c}_{\alpha} \delta t ,t-\delta t), \nonumber \\
   B_y({\bf x},t)&=&\dfrac{1}{\alpha}\sum_{\alpha=1}^9 g_{y\alpha}^{eq}({\bf x}-{\bf c}_{\alpha} \delta t ,t-\delta t),   \\
    B_z({\bf x},t)&=&\dfrac{1}{\alpha}\sum_{\alpha=1}^9 g_{z\alpha}^{eq}({\bf x}-{\bf c}_{\alpha} \delta t ,t-\delta t). \nonumber
\end{eqnarray}
By using the Chapman-Enskog multiscale expansion, it is possible to show that the algorithm \eqref{magnetic field quasi-static acelerated} solves the following equation
\begin{equation}
    \alpha \dfrac{\partial {\bf B} }{\partial t} +\nabla \cdot ( {\bf u} {\bf B}- {\bf B} {\bf u})=\eta \nabla^2 {\bf B},
\end{equation}
with a sufficient number of iterations. 
%This modification essentially decouples the calculus of time derivative from the calculus of the diffusive term in algorithm (\ref{Single step algoritm for magnetic field}), and the introduction of the extra parameter $\alpha$ interferes only in the values of the time derivatives.
This procedure increases the convergence hate by a factor of $1/\alpha$, as we can see in the Figure~\ref{Acelerating convergence}. In this figure, a simulation with a variable pressure gradient given by the formula (\ref{variable force}) is shown. In Figure~\ref{Acelerating convergence}(a), we show a simulations with the algorithm given by (\ref{magnetic field quasi-static acelerated}), where the algorithm for magnetic field is iterated $N=9$ times before every iteration of the algorithm (\ref{single-step algorithm complete}) for the momentum equation. The same experiment is performed by using the algorithm (\ref{Single step algoritm for magnetic field}) with the same number of iterations, i.e., $N=9$, recall that for the algorithm (\ref{Single step algoritm for magnetic field}) the number of iterations is given by (\ref{Number of iterations magnetic field}), which gives $N=18$ for $\eta=3$. We can see that the algorithm (\ref{magnetic field quasi-static acelerated}) converges to the (\ref{ADE for magnetic field}) twice as fast in comparison with (\ref{Single step algoritm for magnetic field}). 

\begin{figure}[H]
		\centering
	\subfigure[]{	\includegraphics[scale=0.6]{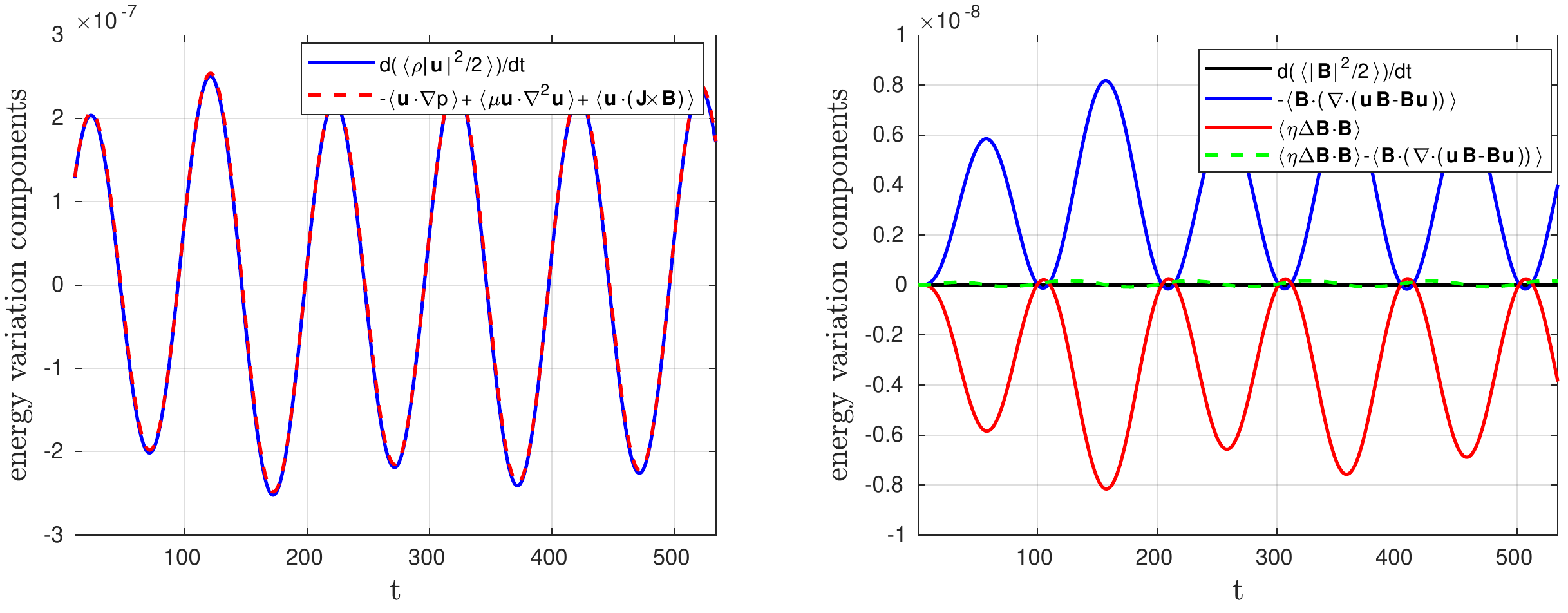}}
	\subfigure[]{\includegraphics[scale=0.52]{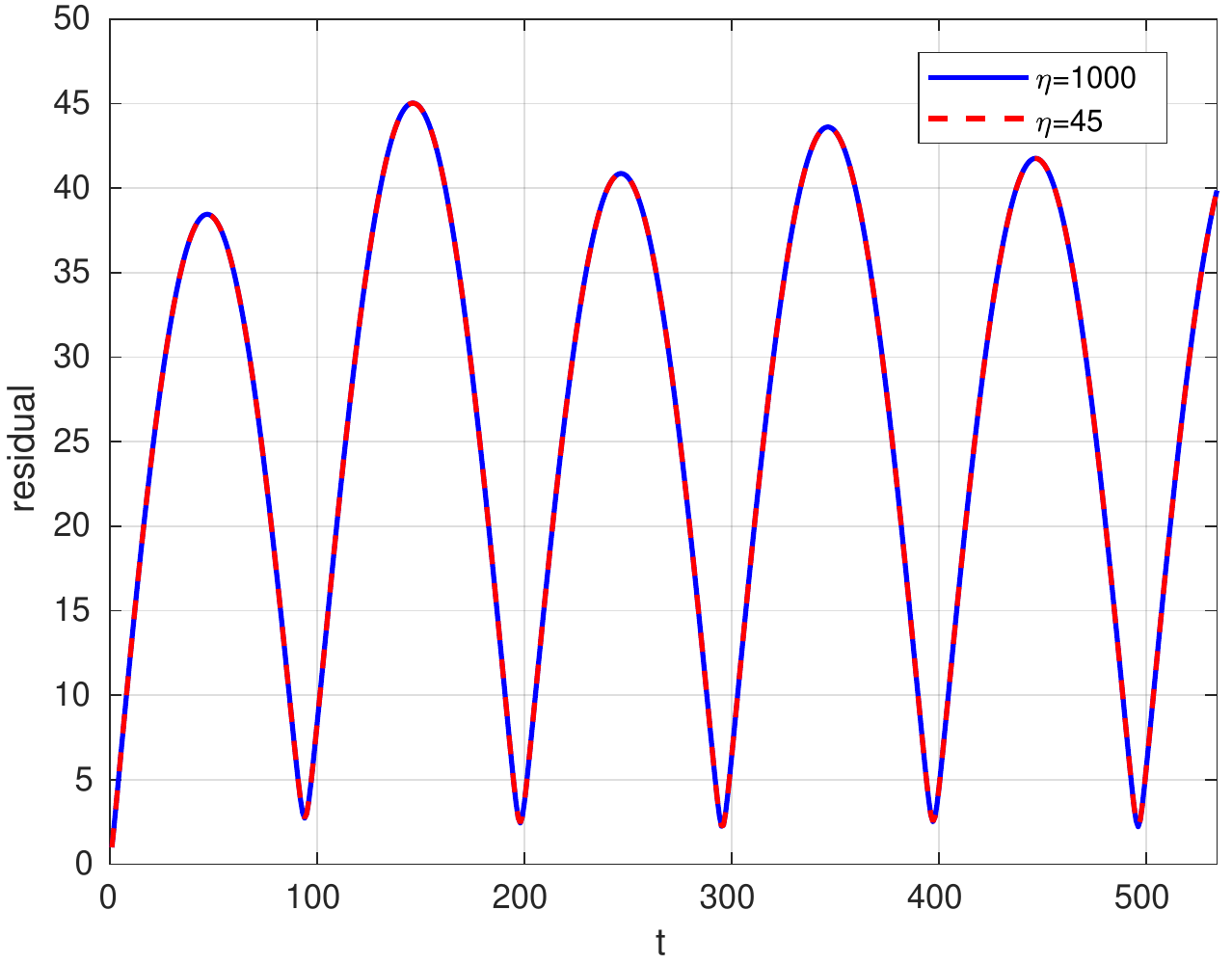}}
		\caption{Simulation of the quasi-static regime in a circular pipe with radius $r=22$, $Ha=18$ and $\nu=0.08$ submitted to $\partial p/\partial x=-1.19\times 10^{-4}\cos(2\pi t/200)$. In the picture (a), the resistivity is set to $\eta=1000$ with $N=270$ iterations of the algorithm \eqref{Single step algoritm for magnetic field} for the magnetic field equations. The analysis of the residuals given by the expression~(\ref{Error l2 decay}) normalized by the initial residual is shown in picture (b), where we compare results for $\eta=45$ and $\eta=1000$.}
		\label{L2 error comparisson}	
	\end{figure}

Values of $\alpha$ smaller than 0.5 can cause instabilities in (\ref{magnetic field quasi-static acelerated}), which limits the application of this procedure with respect to the quasi-static approximation. In Figure~(\ref{Transition to the quasi-static regime})(b), we see that the value of resistivity $\eta=45$ is enough to make the time derivative $\frac{\partial \*B }{\partial t}$ negligible, and the formula~(\ref{Number of iterations magnetic field}) gives $N=270$ as the number of iterations needed to make the the difference $|\eta\nabla^2 \*B-\nabla \cdot (\*u \*B - \*B \*u)|$ to be the same order of the time derivative, thus also negligible. We argue that the same number of iterations is also enough for values of resistivity much bigger then $\eta=45$. In Figure~(\ref{L2 error comparisson}), we compare the simulations with $\eta=45$ and $\eta=1000$ with the same number of iterations $N=270$. In Figure~(\ref{L2 error comparisson})(a), we can see essentially the same results observed in Figure
~(\ref{Transition to the quasi-static regime})(b) in the analysis of the energy budgets.   
	
	In order to calculate more accurately the dependence of the errors with respect to the resistivity with a fixed the number of iterations, we consider the following expression for the residual
	\begin{eqnarray}
	       \left\| \eta \nabla^2 {\bf B} - \nabla \cdot ( {\bf u} {\bf B}- {\bf B} {\bf u})\right\|_{L^2(\Omega)}, \label{Error l2 decay}
	\end{eqnarray}
	%\begin{equation}
	%    \left\| \nabla^2 {\bf B} -\dfrac{1}{\eta} \nabla \cdot ( {\bf u} {\bf B}- {\bf B} {\bf u}) \right\|_{L^2(\Omega)} \ \ \ \textrm{and} \ \ \ \ \left\| \sqrt{\eta} \nabla^2 {\bf B} -\dfrac{1}{\sqrt{\eta}} \nabla \cdot ( {\bf u} {\bf B}- {\bf B} {\bf u}) \right\|_{L^2(\Omega)}
	%\end{equation}
	where the differential operators are calculated by using isotropic finite difference schemes \cite{thampi2013isotropic}. The residual is normalized by the initial residual, i.e., the residual at the first iteration. In Figure~\ref{L2 error comparisson}, we can see that the normalized error (\ref{Error l2 decay}) does not change with the increase of the value of the resistivity, actually the value of the residual \eqref{Error l2 decay}  is the same for $\eta=45$ and $\eta=1000$.
	
  It suggests that for very high values of $\eta$ we do not have to consider a too small value of $\delta t$ (or very high number of iterations) given by the formula~(\ref{Number of iterations magnetic field}), instead we can actually consider an effective number of iterations defined by the number of iterations associated with the smallest value of resistivity that causes a satisfactory damping in the time derivative of the magnetic field, i.e., $|\frac{\partial \*B}{\partial t}|\simeq 0$. An interesting feature of this procedure is that this convergence criterion avoids the need to calculate the residual (which is a non-local procedure) at each time step to guarantee convergence. An alternative procedure is given by solving the equation \eqref{preconditioning previous results} using \eqref{Single step algoritm for magnetic field}, which is equivalent to solve \eqref{preconditioning previous results QS} in the QS regime. Both procedures produces the same results for the experiments of this article.
  
  \section{Effects of non-uniform magnetic fields}\label{Effects of non-uniform magnetic fields}
   
\begin{figure}[h!]
		\centering
		\subfigure[]{
		\includegraphics[scale=0.3]{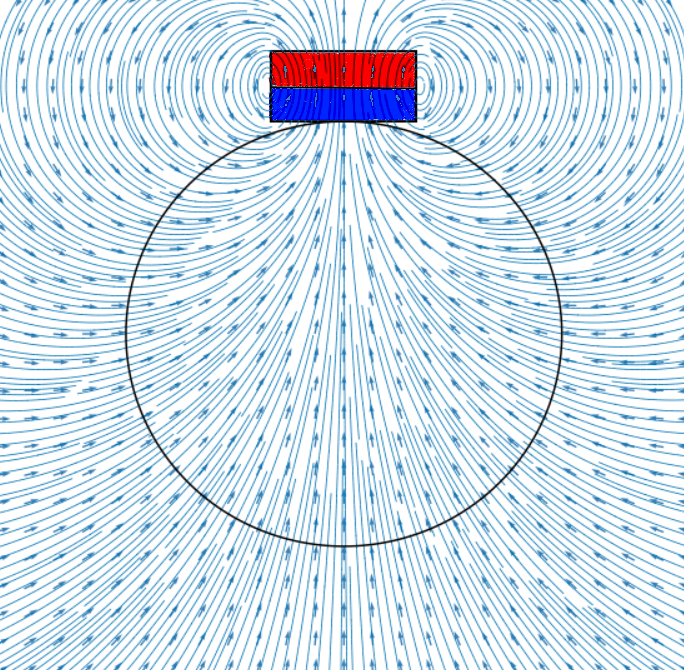}}\hspace{0.5cm}
		\subfigure[]{
			\includegraphics[scale=0.3]{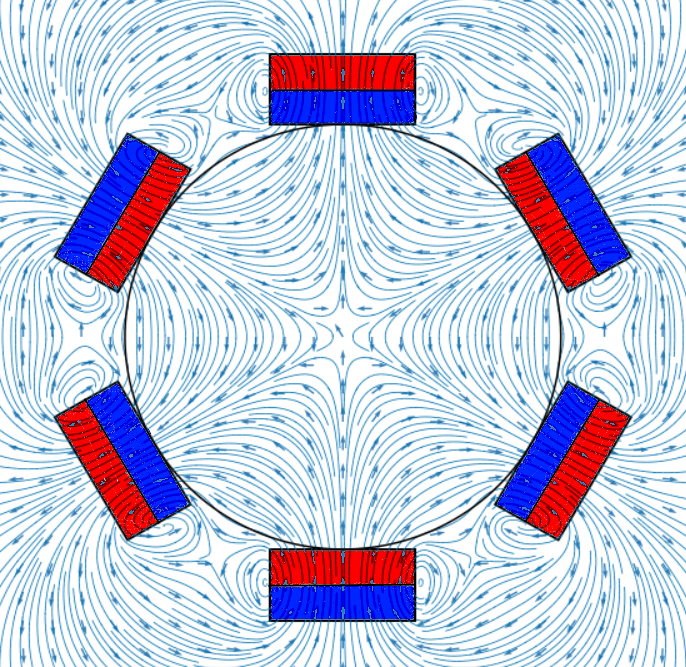}}
		\caption{(a) Magnetic field lines of one magnet generated by the field given by (\ref{x component of the magnetic field}) and (\ref{y component of the magnetic field}) in the yz-plane, where the red and blue colors indicate the north and south poles of the magnets, respectively. In (b), the magnetic field lines generated by a set of six magnets with alternating poles forming a hexagonal structure.}
		\label{One magnetic field lines}	
	\end{figure}

    Most of the LBM simulations of MHD flows only considers the influence of uniform transversal magnetic fields. In this sections, we test the proposed algorithms developed in the previous sections in problems involving an external non-uniform magnetic field, as for example, the field given by 
       \begin{eqnarray}
       \*B_x(x,y,z)&=&0, \label{xx component of the magnetic field} \\
{\bf B}_y(x,y,z)& = & 2 \left[ \arctan\left( \frac{y-L}{z} \right )  - \arctan\left( \frac{y+L}{z} \right) \right] \label{x component of the magnetic field},\\
{\bf B}_z(x,y,z) &=& \log \left ( \frac{(y+L)^2 + z^2}{(y-L)^2 + z^2} \right ),\label{y component of the magnetic field}
\end{eqnarray}
where $(x,y,z)$ is a points in the fluid domain. These fields are
obtained by using the Biot–Savart law~\cite{knaepen2004magnetohydrodynamic}, where $L$ is the width of the slab’s rectangular cross section, which we assume to have
aspect ratio 2. We consider $L=R/6$.
The magnetic field lines generated  by~(\ref{x component of the magnetic field}) and (\ref{y component of the magnetic field}) in the yz-plane are shown in Figure~\ref{One magnetic field lines}(a); and in Figure~\ref{One magnetic field lines}(b) we shows the field lines of a combination of six magnets with alternating poles, where the magnetic field of each magnet can be mapped into the field given by \eqref{xx component of the magnetic field}, (\ref{x component of the magnetic field}) and (\ref{y component of the magnetic field}) by considering compositions with rotations and translations.

\begin{figure}[H]
		\centering
		\subfigure[]{
\includegraphics[scale=0.79]{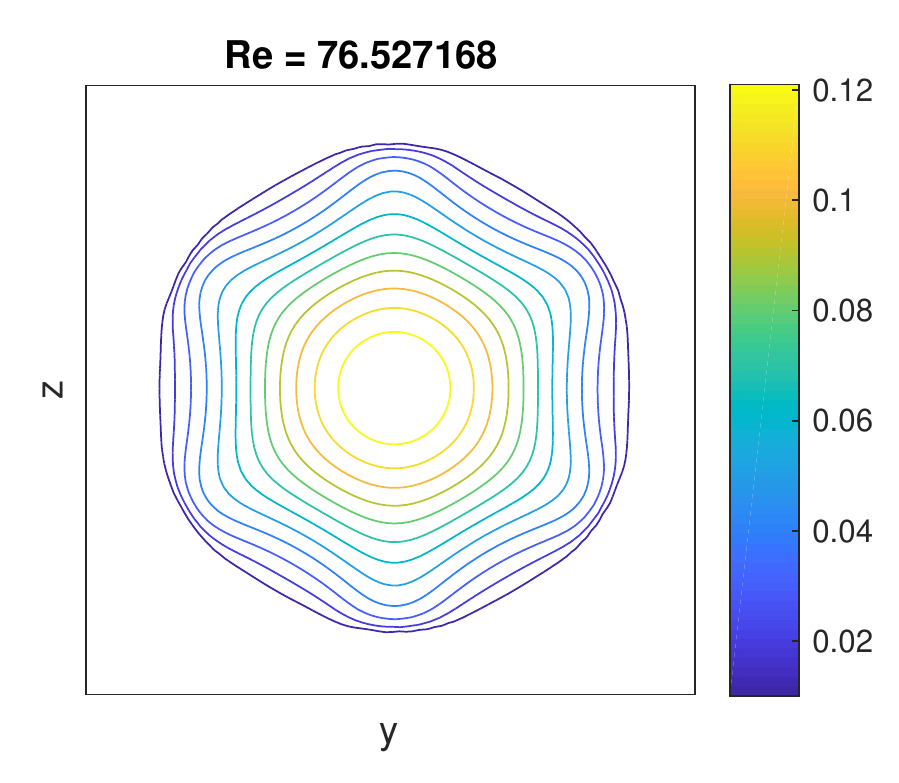}}\hspace{0.5cm}
  		\subfigure[]{
\includegraphics[scale=0.75]{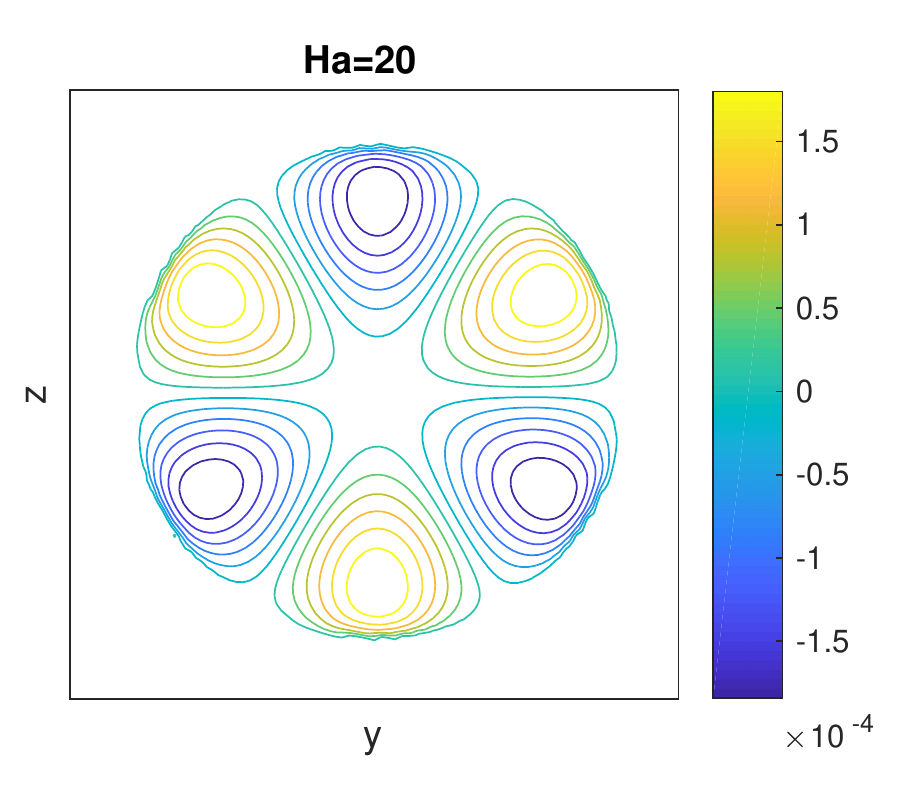}}
		\subfigure[]{
	\includegraphics[scale=0.6]{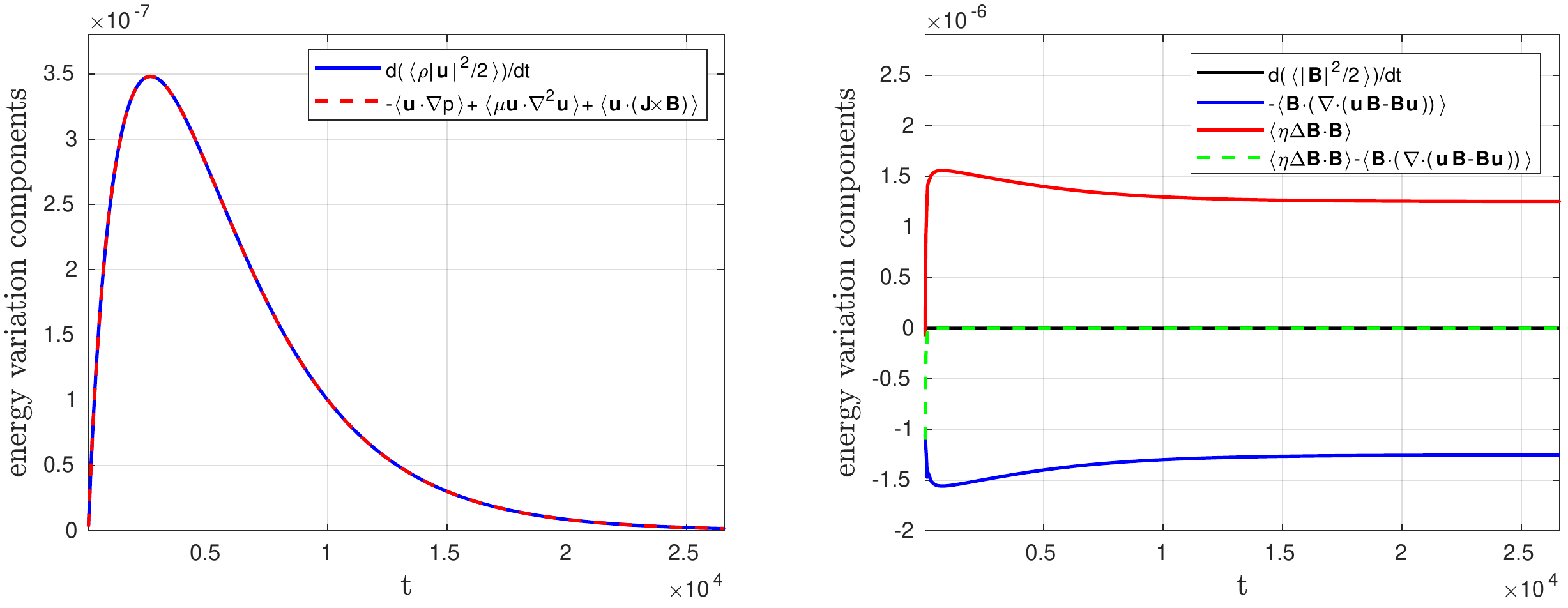}}
		\caption{Simulation of a MHD flows with the six magnets configuration in the quasi-static regime with $\eta=1000$, $Ha=20$, $\nu=0.04$ and pipe radius $r=38.5$. A constant body force with $\frac{\partial p}{\partial x}=-2.16\times10^{-5}$ is applied. The algorithm for the magnetic field (\ref{Single step algoritm for magnetic field}) is iterated $N=12$ times before every iteration of the algorithm for the velocity field (\ref{Single step forcing term introdution}). The simulation is performed until the stationary solution is obtained. In (a) and (b) we show some level curves for the velocity and magnetic fields, respectively. We can see that the velocity and magnetic field profiles
  verify the expected symmetries associated with the system (\ref{Navier-Stokes equations for QS}-\ref{zero divergence magnetic field}). In (c) we show the verification of the energy balance equations given by  (\ref{Energy balance equations}).}
		\label{Energy balance non-uniform MHD}	
	\end{figure}

	In Figure~(\ref{Energy balance non-uniform MHD}), we show a simulation of a MHD flow in a circular pipe based on the schematic representation shown in Figure~\ref{One magnetic field lines}(b), with viscosity $\nu=0.04$, resistivity $\eta=1000$, Hartman number  $Ha=20$ and pipe radius $r=40$ in a computational grid with size  $n_x \times n_y \times n_z = 5\times 83 \times 83$ . Periodic boundary conditions are considered in the streamwise direction and a constant body force $\frac{\partial p}{\partial x}=-2.16\times10^{-5}$ is imposed. In the Figures~\ref{Energy balance non-uniform MHD}(a) and \ref{Energy balance non-uniform MHD}(b), we can observe the contour lines for the velocity and magnetic fields showing the expected symmetry associated with the magnetic field configuration presented in Figure~\ref{One magnetic field lines}(b). The respective verification of the energy balance \eqref{Energy balance equations} is shown in Figure~\ref{Energy balance non-uniform MHD}(c). The modification of the equilibrium distributions in order to implement the divergence of the Maxwell stress tensor~\cite{de2021one}, rather than the direct implementation of the Lorentz force, has not shown stable results for the cases involving non-uniform magnetic fields, indicating that for the algorithms presented in this article, the forcing term approach given by \eqref{Single-step algorithm with lattice discrete effect} is more suitable procedure.

  \section{Simulations with magnetic Prandtl number $Pr_m>1$}\label{Prandtl bigger than 1}
  
\begin{figure}[h!]
		\centering
	\subfigure[]{	\includegraphics[scale=0.6]{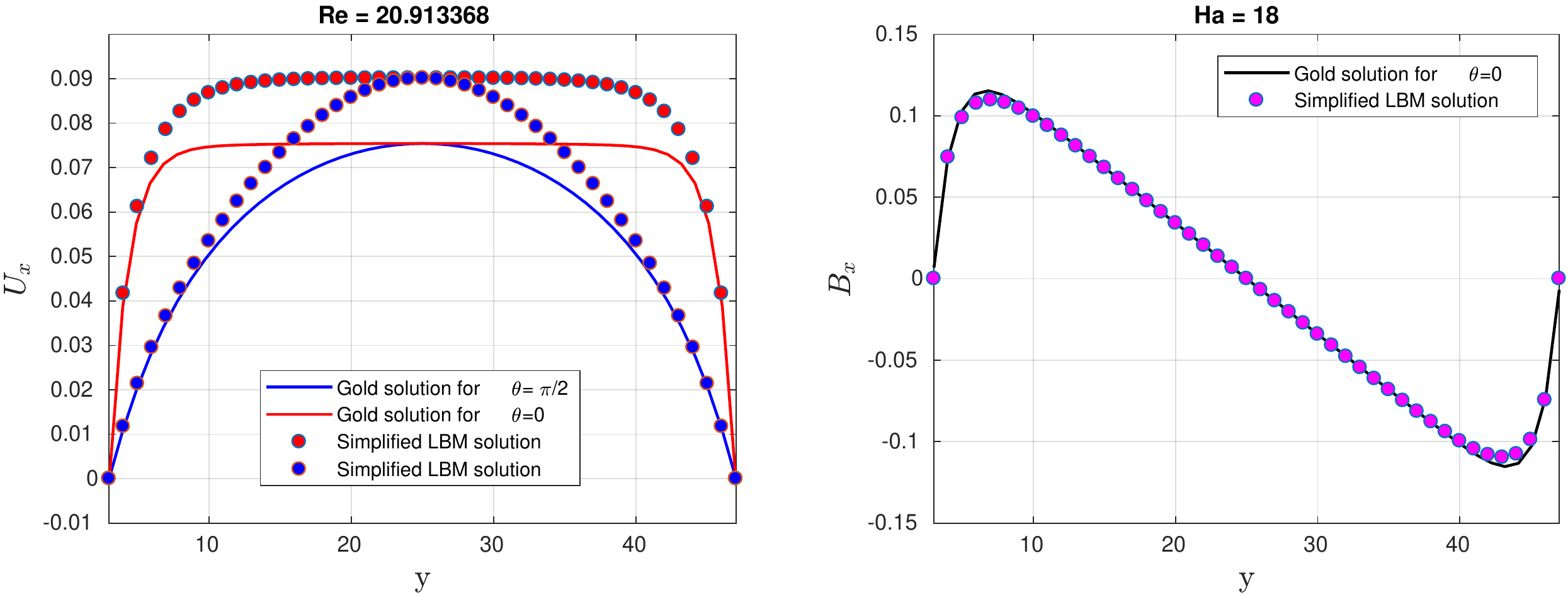}}\hspace{1cm}
	\subfigure[]{	\includegraphics[scale=0.6]{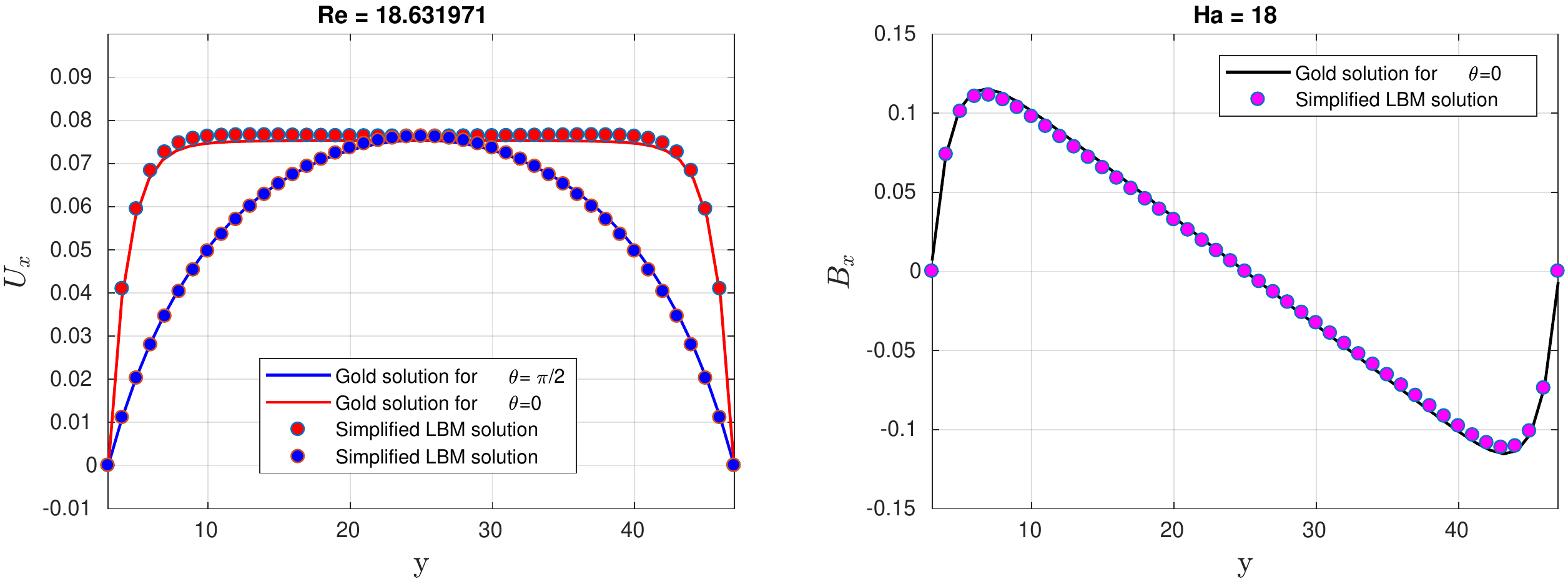}}
		\caption{MHD pipe flow simulations with $\nu=0.08$, $\eta=0.02$, $Ha=18$ and pipe radius $r=22$ with $\gamma=1$ in picture (a), and $\gamma=4$ in picture (b). A constant body force with $\frac{\partial p}{\partial x}=-2.38\times10^{-4}$ is applied. It is possible to observe a significant loss of spatial accuracy in (a), which is restored by using the rescaled simplified single-step LBM solution obtained by using \eqref{adaptative time step delta t} and \eqref{adaptative time step lattice velocities}, as we can see in picture (b).}
		\label{Anti-prefactors profiles}	
	\end{figure}

  In all of the previous discussions, we concentrate our analysis in regimes with $Pr_m\leq1$. In this section, we analyse the results of the single-step simplified algorithms proposed in this article for the case $Pr_m>1$. This regime usually requires more accuracy of the numerical methods in space and time. The few LBM results in the LBM literature~\cite{de2018advanced,de2022vortex} about this regime are performed up to $Pr_m=2$ by using more robust numerical schemes, such as the central-moments-based LBM in simulations with flat boundaries. In Figure~\ref{Anti-prefactors profiles}(a), we can see a significant mismatch between numerical and analytical solution  by using~(\ref{Single step forcing term introdution}) with~(\ref{Single step algoritm for magnetic field}) in a simulation with $Pr_m=4$, despite the improvements introduced in the previous sections. 

 In order to solve this problem we consider a strategy based in the introduction of a smaller time steps. Most of the simplified LBM methods are constructed considering $\delta t=\delta x$, which restricts the possibilities of the changes of $\delta t$ to some particular grid configurations. In order to avoid this limitation, we consider a set of rescaled variables, indicated by overlines, associated with an extra parameter $\gamma$ verifying 
  \begin{equation}\label{adaptative time step delta t}
      \overline{\delta t} = \dfrac{\delta t}{\gamma},
  \end{equation}
  with rescaled lattice velocities $\overline{\*c_i}$ and speed of sound $\overline{c}_s$ verifying 
  \begin{equation}\label{adaptative time step lattice velocities}
      \overline{\*c_i} = \gamma \*c_i  \ \ \ \textrm{and} \ \ \  \overline{c}_s = \dfrac{\gamma c }{\sqrt{3}}= \gamma c_s,
  \end{equation}
   which means that we are keeping the computational grid unchanged, i.e., $\delta x=\overline{\delta x}$. Assuming that the density is not affected by the transformations, i.e., $\overline{\rho} \simeq \rho $, it turns out that the substitution of scaling relationships \eqref{adaptative time step delta t} and \eqref{adaptative time step lattice velocities} in the demonstration of the single-step algorithm~\eqref{single-step algorithm complete} is equivalent to consider the original algorithm (where $\delta x=\delta t$) with the rescaled equilibrium distributions
  		\begin{eqnarray}
		f_i^{eq}(x,t)
		&=&w_i\rho\left[1+\dfrac{\mathbf{c}_i\cdot \mathbf{u}}{c_s^2}
		+\dfrac{1}{\gamma}\left(\dfrac{(\mathbf{c}_i\cdot \mathbf{u})^2}{2 c_s^4}
		-\dfrac{\mathbf{u} \cdot \mathbf{u}}{2c_s^2} \right) \right],\\
		g_{ix}^{eq}(x,t) &=& w_i\left[ B_x +\dfrac{1}{\gamma}\left( \dfrac{c_{iy}}{c_s^2} (u_y B_x - u_x B_y)+ \dfrac{c_{iz}}{c_s^2} ( u_z B_x - u_x B_z) )  \right) \right],\\
		g_{iy}^{eq}(x,t) &=& w_i \left[ B_y +\dfrac{1}{\gamma} \left( \dfrac{c_{ix}}{c_s^2} (u_x B_y - u_y B_x)+ \dfrac{c_{iz}}{c_s^2} ( u_z B_y - u_y B_z) ) \right)  \right],\\
		g_{iz}^{eq}(x,t) &=& w_i \left[ B_z +\dfrac{1}{\gamma} \left( \dfrac{c_{ix}}{c_s^2} (u_x B_z - u_z B_x)+ \dfrac{c_{iy}}{c_s^2} ( u_y B_z - u_z B_y) )  \right) \right],
		\end{eqnarray}
  and rescaled relaxation times given by
  \begin{equation}\label{rescaled relaxation time}
      \tau =  \dfrac{\nu}{\overline{c}_s^2 \overline{\delta t} }+\dfrac{1}{2} = \dfrac{\nu}{\gamma c_s^2 \delta t }+\dfrac{1}{2},
  \end{equation}
  and
  \begin{equation}
      \tau_m = \dfrac{\eta}{ \overline{c}_s^2\overline{\delta t} }+\dfrac{1}{2} = \dfrac{\eta}{\gamma c_s^2 \delta t}+\dfrac{1}{2},
  \end{equation}
    which are defined in such a way to keep the viscosities and resistivities unchanged by the transformations \eqref{adaptative time step delta t} and \eqref{adaptative time step lattice velocities}. 
    If we consider the inclusion of the FGS forcing term \eqref{FGS forcing term algorithm} in the simplified algorithm, then the introduction of \eqref{adaptative time step lattice velocities} also leads to
    \begin{eqnarray}
	F_i = \left(1-\dfrac{1}{2\tau}\right)w_i \left[ \dfrac{(\*c_i-\frac{\*u}{\gamma})}{\gamma c_s^2}+\dfrac{(\*c_i \cdot \*u )}{ \gamma^2 c_s^4} \*c_i \right] \cdot \*F_{ext}.
\end{eqnarray}
with $\tau$ given by \eqref{rescaled relaxation time}.
 All of these modifications provide essentially the same result as those obtained using the preconditioning procedures described in \cite{premnath2009steady,guo2004preconditioned,izquierdo2008preconditioned,turkel1999preconditioning}. The same equations can also be found by using the strategy of the adaptive time step (ATS) developed in~\cite{horstmann2022consistent}, with the exception of the treatment of the nonequlibrium terms. 
    
 If we consider we consider $\gamma>1$,  we essentially decrease of the effective time step by a factor of $1/\gamma$. Consequently, we obtain a significant improvement of the accuracy with minimum changes in the original single-step algorithm. Naturally, $\gamma$ cannot be changed arbitrarily, if $\gamma$ is too small, some transient phenomena with typical small time scales may be missed, and if $\gamma$ is too large, simulations may have an excessively slow convergence rate with some possible loss of accuracy, due to the fact that the scalings can make the relaxation times to close to the value $0.5$ if $\gamma\gg 1$.

  \begin{figure}[h!]
		\centering
		\subfigure[]{
		\includegraphics[scale=0.6]{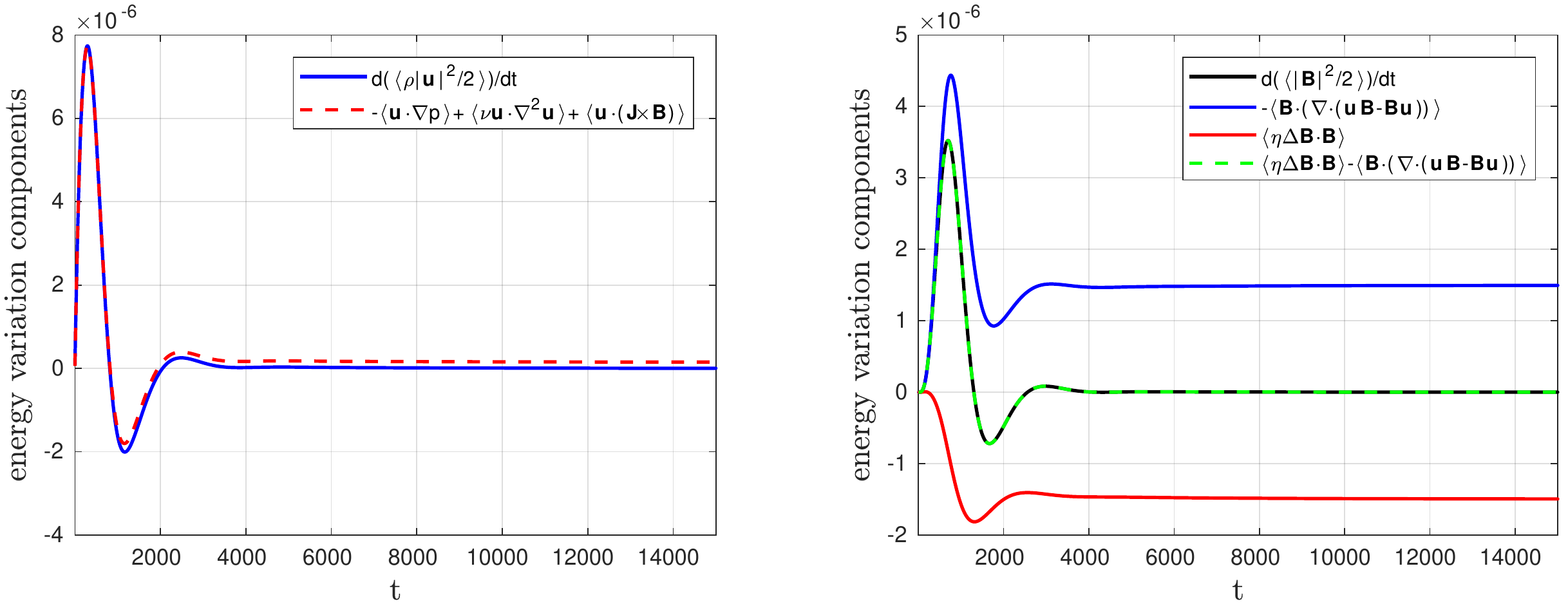}}\hspace{1cm}
		\subfigure[]{
		\includegraphics[scale=0.6]{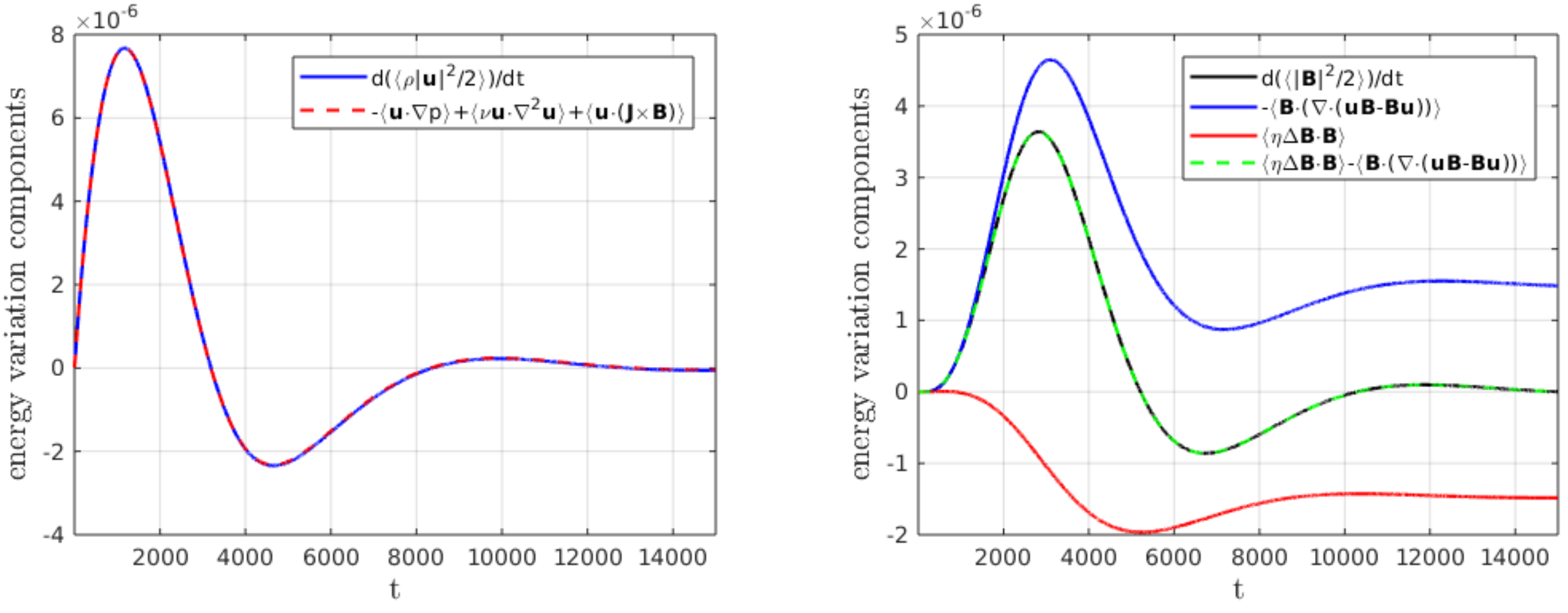}}
		\caption{MHD pipe flow simulation with $\nu=0.08$, $\eta=0.02$, $Ha=18$ and pipe radius $r=22$. A constant body force with $\frac{\partial p}{\partial x}=-2.38\times10^{-4}$ is applied. In picture (a), we show results of the analysis of the energy balance (\ref{Energy balance equations}) for $\gamma=1$, and  in picture (b), the results for $\gamma=4$. By analysing the transient part in the energy budgets, we can observe an increase of the accuracy in time by using the rescaled simplified single-step LBM solution obtained by introducing \eqref{adaptative time step delta t} and \eqref{adaptative time step lattice velocities}.}
		\label{Anti-preconditioning energy}	
	\end{figure}

	In the Figures~\ref{Anti-prefactors profiles} and~\ref{Anti-preconditioning energy}, we performed some MHD pipe flow simulations with $\nu=0.08$, $\eta=0.02$, $Ha=18$ and pipe radius $r=22$. A constant body force with $\frac{\partial p}{\partial x}=-2.38\times10^{-4}$ is applied.  The computational grid size considered is $n_x \times n_y \times n_z=5 \times 50 \times 50$.   In Figures~\ref{Anti-prefactors profiles}(b) and~\ref{Anti-preconditioning energy}(b), we show the velocity and magnetic field statistics associated with the scaling $\gamma=4$; and in  Figures~\ref{Anti-prefactors profiles}(a) and~\ref{Anti-preconditioning energy}(a), we present the statistics generated by using $\gamma=1$. In the Figure~\ref{Anti-preconditioning energy}, it is possible to see that the solutions were essentially rescaled in time by a factor of $\gamma$. Not only that, we can also observe a significant improvement of the accuracy in space (verification of the Gold's solutions) and time (correct verification of the energy balance).

  \section{Conclusions}\label{Conclusions and Perspectives}

  In this article, we provide a set of extensions and improvements in a class of simplified LBM algorithms with the objective to simulate MHD flows with very small magnetic Reynolds numbers in pipe flows. We also introduce a immersed boundary method able to accurately include the effects curved insulating walls in the MHD equations and whose accuracy is not significantly dependent on the values of the relaxation times. Improvements in the implementation of forcing term allows an accurate and stable implementation of variable forcing terms, showing good results even in the presence of strongly non-uniform magnetic fields.
  With this set of improvements, in the present work we provide
  a completely local and explicit LBM framework for simulations of the quasi-static approximation in pipe flows, with a good potential for simulations involving more complex geometries.
  
  By considering an adaptive time step strategy, we were able to increase the precision of the single-step LBM algorithm in space and time with minimal changes in the general form of the algorithm, extending the applicability of the method to some regimes up to $Pr_m=4$, which have not yet been analyzed in the LBM literature. It is also important to mention that results introduced in this article can be extend as well to some other simplified lattice Boltzmann models~\cite{chen2017simplified,inamuro2002lattice,shu2014development,zhou2020macroscopic}.
  
  As future works and suggestions, further verification of the proposed methods for turbulent flows and extensions for the cases involving conducting curved walls are natural future directions for this research, as well as systematic comparisons with similar solutions provided by other numerical methods. Also, the use of of more robust LBM schemes such as MRT (multiple-relaxation-time) and central-moments-based schemes for the magnetic field equations can be  interesting options towards the same objectives of this article, with some possible improvements in terms of accuracy~\cite{magacho2022double}.  

\section*{Acknowledgements}

The authors acknowledge the support given to this work by the project entitled \quotes{Experimental study
of inorganic fouling in sand containment systems}, established by agreement between the
COPPETEC Foundation (COPPE/UFRJ) and the oil company Petróleo Brasileiro S.A. (Petrobras), with the project of number 21.389. They also thank the Interdisciplinary Center of Fluid Dynamics (NIDF) at UFRJ, which was of great help in terms of people, infrastructure and resources for the research presented in this article.

%%%%%%% Comment 

\bibliographystyle{elsarticle-num}
\bibliography{Bibliografia}

\end{document}